\documentclass{article}
\usepackage[a4paper,margin=1in]{geometry}
\usepackage{graphicx} 
\usepackage{epstopdf} 
\usepackage{qcircuit}
\usepackage{subfiles}
\usepackage{xcolor}
\usepackage{booktabs}
\usepackage{enumerate}
\usepackage{float}
\usepackage{anyfontsize}
\usepackage{caption}
\usepackage{tabularx}
\usepackage{array}
\usepackage{changepage}
\usepackage[style=ieee]{biblatex}
\usepackage[colorlinks=true, allcolors=blue]{hyperref}
\usepackage{orcidlink}

\usepackage[hang,flushmargin,stable]{footmisc}
\usepackage{chngcntr}
\counterwithout{footnote}{page}

\usepackage{scrextend}
\deffootnote[1.9em]{1.9em}{1em}{\textsuperscript{\thefootnotemark}\,}

\makeatletter
\let\@fnsymbol\@arabic 

\renewcommand{\thanks}[1]{%
  \refstepcounter{footnote}%
  \footnotemark[\value{footnote}]%
  \protected@xdef\@thanks{%
    \@thanks
    \protect\footnotetext[\the\value{footnote}]{\protect\parbox[t]{0.9\linewidth}{#1}}%
  }%
}
\makeatother

\providecommand{\keywords}[1]{\small\textbf{\textit{Keywords---}} #1}

\title{SoK: A Systematic Review of Malware Ontologies and Taxonomies and Implications for the Quantum Era}
\date{August 2025}

\author{%
Dehinde Molade~\orcidlink{0009-0007-2551-9342}\thanks{Dehinde Molade, Department of Computer and Information Science, University of South Australia (UniSA), Building W, Mawson Lakes SA 5095, Australia. \textit{Corresponding author.} Email: \href{mailto:dehinde.molade@mymail.unisa.edu.au}{dehinde.molade@mymail.unisa.edu.au}.}\\
Dave Ormrod~\orcidlink{0000-0002-0599-8362}\thanks{Dave Ormrod, CTO and Co-Founder, Cygence, Australia (Industry Supervisor).}\\
Mamello Thinyane~\orcidlink{0000-0001-5900-0601}\thanks{Associate Professor Mamello Thinyane, Optus Chair in Cybersecurity UniSA STEM, Mawson Lakes SA 5095, Australia.}\\
Nalin Arachchilage~\orcidlink{0000-0002-0059-0376}\thanks{Associate Professor Nalin Arachchilage, in Cyber Security, RMIT University, School of Computing Technologies, Melbourne, Australia.}\\
Jill Slay~\orcidlink{0000-0002-2352-8815}\thanks{Professor Jill Slay, SmartSat Chair in Cybersecurity UniSA STEM, Mawson Lakes SA 5095, Australia.}
}

\begin{document}
\setcounter{footnote}{0}
\maketitle
\begin{abstract}
\normalsize
The threat of quantum malware is real and a growing security concern that will have catastrophic scientific and technological impacts, if not addressed early. If weaponised or exploited especially by the wrong hands, malware will undermine highly sophisticated critical systems supported by next-generation quantum architectures, for example, in defence, communications, energy, and space. This paper explores the fundamental nature and implications of quantum malware to enable the future development of appropriate mitigations and defences, thereby protecting critical infrastructure. By conducting a systematic literature review (SLR) that draws on knowledge frameworks such as ontologies and taxonomies to explore malware, this provides insights into how malicious behaviours can be translated into attacks on quantum technologies, thereby providing a lens to analyse the severity of malware against quantum technologies. This study employs the European Competency Framework for Quantum Technologies (CFQT) as a guide to map malware behaviour to several competency layers, creating a foundation in this emerging field. 
\end{abstract}

{\color{white}-}\\
{\color{white}-}\\

\section{Introduction}
\normalsize
Quantum mechanical systems, perceived as a herald and vanguard of novel simulation, unbreakable encryption, and digital autonomy, perhaps someday, will become the very instruments of precision-targeted, well-aimed campaigns towards state-scale collapse. Quantum decryption, synthetic biology, bioweapon engineering, disruption of nuclear safeguards \supercite{farley2021quantum} or protocols, weaponised sensing, quantum blackmail, and nullification of nuclear deterrence \supercite{gamberini2021quantum} may soon evolve to represent perhaps one of the greatest threats to geopolitical and societal stability, creating cascading effects capable of increased tension and conflict, civil unrest, endangering modern civilisation. This work aims to expose these looming spectres of the deployment of digital weapons, particularly malware, against critical quantum infrastructures. As described in the literature, when a malicious actor compromises classical infrastructure systems, such as power grids, water supplies, or communication networks, the resulting impact can be severe and far-reaching. With advancement in quantum especially being embedded into critical national infrastructure to bolster security, sensing and computational dominance \supercite{karol2015quantum, alghamdi2023exploring, raymer2019us, monroe2019us, brinkley2022future}, this could be accelerating the arrival of unprecedented catastrophic security events through the convergence of quantum innovation with malicious intent, ushering in a new era of cyber kinetic warfare, where the consequences are not digital, but existential.

Quantum key distribution (QKD) will enhance the secure distribution of cryptographic keys protecting sensitive communication against future quantum attacks and eavesdropping \supercite{gisin2002quantum, pirandola2020advances, bennett1992quantum}, while quantum sensors (e.g., gravimeters, magnetometers and accelerometers) will provide ultra-precision monitoring and geolocation ability of physical systems such as power grids, pipelines, mineral and oil. It will also enhance transportation networks by improving logistics, route accuracy, and enabling reliable geolocation tracking independent of GPS satellites \supercite{degen2017quantum, kantsepolsky2023exploring}. Quantum computers will support information processing for complex large tasks and applications, such as novel chemical synthesis to discover new drugs \supercite{santagati2024drug, cao2018potential, blunt2022perspective} and materials\supercite{louie2021discovering, bauer2020quantum}, optimise energy distribution \supercite{ajagekar2019quantum}, advanced manufacturing \supercite{bova2021commercial}, and analyse large data sets more effectively to detect anomalies in financial transactions \supercite{orus2019quantum}. Despite technological advancements in quantum technologies, it will be implemented alongside classical systems in a hybrid fashion, with quantum components managing tasks that require high precision, security, or processing power, and classical systems functioning as the control and interface for operating quantum hardware, thereby strengthening the resilience and intelligence of critical infrastructures. The hybrid architecture combining quantum and classical systems is likely to be the standard environment attackers and defenders will face for the foreseeable future.

Unfortunately, quantum systems and infrastructures used for computational, communication and sensing services will likely be disrupted by cybersecurity attacks \supercite{NQCO2022Cybersecurity, ghosh2023primer, baseri2024cybersecurity, johnson2019quantum, satoh2021attacking, wu2006quantum}. However, due to the quantum mechanical rule known as the no-cloning theorem, generating an indistinguishable clone of an unknown quantum state is non-viable \supercite{wootters1982single, milonni1982photons}. Therefore, it is inherently difficult to reverse engineer quantum states. However, through the dependency of classical systems via integrated circuits (e.g., CMOS), along with software libraries, and compilers \supercite{almudever2021quantum} to control, measure, manage, and support quantum states, i.e., Qubits, may result in potential weaknesses and vulnerabilities. There are dangers associated with quantum technology stacks that comprise both classical hardware and software, which often depend heavily on third-party components, many of which may only be partially trusted or entirely unreliable. Network-level threats, such as multi-tenant cloud-based computation attacks and unauthorised network and cloud access at the network layer, could also pose as a security risk. \supercite{ghosh2023primer} While current classical systems remain cloneable and hackable, they are also particularly susceptible to malware infections \supercite{kramer2010general, milovsevic2013history, alenezi2020evolution} which can be delivered and controlled remotely by adversaries, leaving gaps for security and privacy concerns toward a fully secure quantum architecture \supercite{ghosh2023primer}.

This study aims to examine the nature of malware-related threats that may emerge as quantum architectures are integrated into critical national infrastructure. It investigates known classical malware behaviours, codified using ontological and taxonomical frameworks, and correlates them with potential security vulnerabilities and weaknesses typically identified in various quantum components. This research leverages the European Competence Framework for Quantum Technologies (CFQT) \supercite{cfqt2025, greinert2025extending} as a foundation for modelling quantum technology environments based on its broad domain coverage, pan-European reference standard, and facilitation of cross-domain analysis, which allows the formal mapping of malware behaviours to specific layers of a quantum stack. The study analyses the wide range of security risks posed by malware threats, highlighting the potential impact on quantum technologies embedded within critical and high-value infrastructure. Furthermore, it considers the speculative yet plausible future consequences of classical and quantum native malware adapting to target quantum systems, such as the quantum Internet, network nodes, quantum applications, and other underlying architectures.

Finally, considering how ontologies \supercite{uschold1996ontologies, noy2001ontology, guber1993translational, mcguinness2003ontologies} and taxonomies \supercite{nickerson2013method, sowa1999knowledge, bailey1994typologies} have supported knowledge formalisation and modelling across various fields by providing a well standardised, formal structure and shared language across complex domains aiding the reliability for consistent knowledge communication, understanding, interoperability, reuse and reasoning, the study investigates the potential use of ontologies and taxonomies for the knowledge formalisation of malware in both the classical and quantum domain and proposes the development of a quantum malware ontology, drawing inspiration from existing classifications of classical malware, to structure and contextualise these emerging threats.

The paper begins with Section 2, which explains ontologies and taxonomies, malware, and explores the implications for quantum. Section 3 presents the systematic review performed to answer the following key questions.  
\begin{enumerate}
    \item[\textbf{Q1:}] How can ontologies help us define and understand the fundamental nature of malware?
    \item[\textbf{Q2:}] How do taxonomies help us systematically classify malware behaviours, lifecycle, and categories?
\end{enumerate}

Then Section 4 discusses the findings of the review. The final section synthesises the findings of the review. It explores the security implications of future quantum interconnected systems, particularly the identification and mapping of the risks to quantum components within the CFQT based on insights from the review.

\section{Background and related works}
This research is primarily framed across three key lines of inquiry: the first explores the role of ontologies and taxonomies in codifying knowledge about malware, the second interrogates the current state-of-the-art on classical malware, and the last presents a discussion on the implications of malware in the quantum era.
\normalsize

\subsection{Ontologies and taxonomies}
Ontologies and taxonomies have significantly advanced human understanding by providing formalised language and definitions that facilitate clear communication and collaboration between researchers and industry professionals \supercite{Behaviouralontology}.

As a philosophical discipline, ontology investigates the basic structure and essence of reality \supercite{guarino2009ontology}. As a science, it is concerned with the reality of using formal logic, often defined as a tuple \supercite{navarro2018leveraging} consisting of classes, subclasses, and properties to describe behaviours, events, processes, and relationships that exist between them and the reality or nature of being to provide a well-informed conceptual description of a knowledge area or entity that helps facilitate knowledge sharing and reuse of information \supercite{ding2019ontology}.

As a formalised way of representing knowledge concepts and relationships between various concepts or ideas, it has been used in various domains, including information engineering, Artificial Intelligence (AI), Web semantics, and information security. Often described as human and machine interpretable, ontologies \supercite{Behaviouralontology} help serve as a common language for human-machine interaction and reasoning, describing hierarchies, classes, inheritance, and relationships based on accepted truths (axioms). Ontologies can also integrate the usage of queries and rules to make better decisions in real-world scenarios.

With designs around main classes that define the domain, sub-classes for specific details and super-classes for more general information. The class relationships determine how objects interact, while the instances represent unique examples of classes made up of property values of data and property connections of objects \supercite{rastogi2020malont}. Ontologies have also been described as vital for representing complex information and systems using lexical concepts with definitions and machine-interpretable semantics \supercite{huang2014it2fs} even when dealing with uncertainty. The limitations of conventional ontologies in dealing with uncertain or indefinite knowledge have led to the emergence of fuzzy ontologies that are better suited for intelligent design, enhancing real-world applicability, and improving transparency and reliability. FML (Fuzzy Markup Language) is an example of a novel ontological language to help model imprecise or fuzzy systems, improving realistic expectations. 

Modelling malware behaviours may be implicit; hence, adopting an FML-based ontological agent can aid malware behaviour analysis in designing intelligent decision-making systems useful for anti-malware protection \supercite{huang2011applying}. Security ontologies facilitate the development of threat classifications, risk evaluations, and architectural software modelling frameworks \supercite{navarro2018leveraging}. It can also be useful for identifying and preventing the spread of cyber attacks by providing a deeper understanding of the cybersecurity domain \supercite{UnifiedOntology} and modelling relevant concepts related to malware behaviours, classes, variants, and computer system units \supercite{ding2019ontology}.

\begin{figure}[H]
    \centering
    \includegraphics[width=10cm, height=7cm]{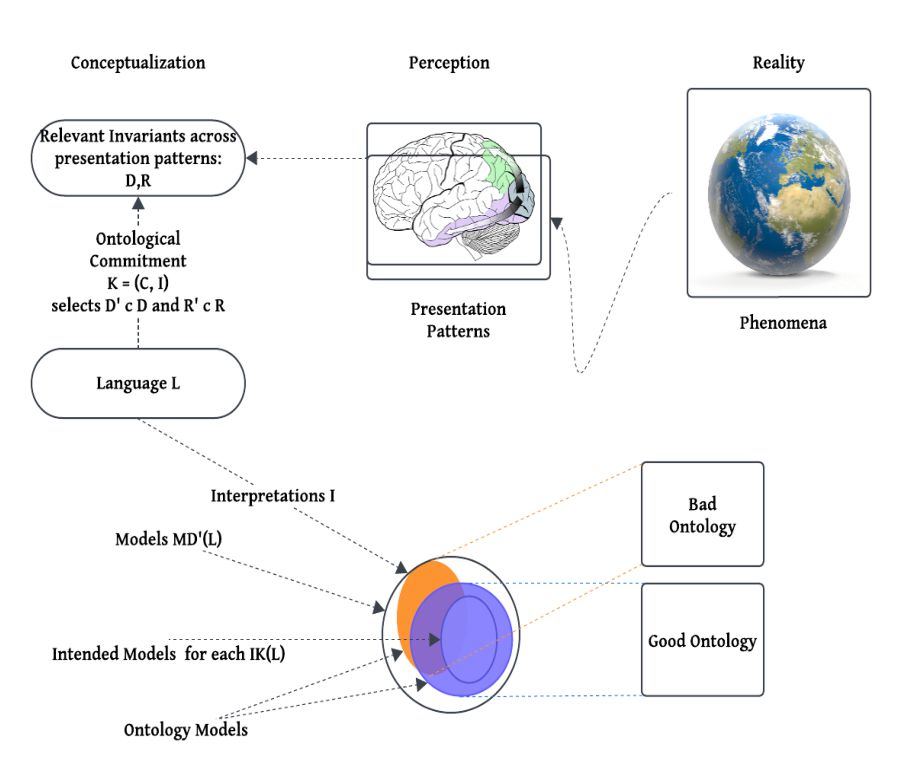} \
    \caption{How ontology improves perception of real-world phenomena. Adapted from \supercite{guarino2009ontology}.}
    \label{fig:ontologyframe}
\end{figure}

An effective ontology helps us perceive and understand reality more accurately by providing clear, consistent definitions and relationships between concepts. However, ontologies can also lead to misunderstandings and misinterpretations that affect how the world is understood and shaped. This can be a significant issue for ontologies that underpin AI systems, where human contextualisation may not be present to offset a misinterpretation within the system. Figure~\ref{fig:ontologyframe} illustrates the relationships between these key concepts. It shows how phenomena in reality are perceived, abstracted into concepts, described using language, and represented by intended models. The ontology serves as the framework that binds all these elements together, ensuring that human perceptions and descriptions are consistent and accurate \supercite{guarino2009ontology}.

Taxonomies can be defined as a conceptually or empirically derived way of making classifications or categories \supercite{nickerson2013method}. Originating in biological sciences \supercite{padial2010integrative} as a systematic approach to grouping living organisms, its foundation was established by Carl Linnaeus in the 18th century \supercite{godfray2002challenges}. Linnaeus introduced the binomial nomenclature or Linnaean system of classification to hierarchically categorise living organisms into six taxonomic ranks \textit{"Regnum (Kingdom), Classis (Class), Ordo (Order), Genus (Genus), Species (Species) and Varietas (Variety)"} \cite[p.~126]{de1997linnaean} attributed within each of the two primary kingdoms (Animalia and Plantae). Its universality, scalability, and interoperability allowed its evolution beyond biological theories. The unique ability of the system to capture complex information in a structured, clear, and reusable format has also enabled its application to various disciplines, including information science \supercite{nickerson2013method}, enterprise knowledge management\supercite{lambe2014organising}, cognitive education\supercite{bloom1956taxonomy}, computer science, artificial intelligence, business, and, particularly, cybersecurity \supercite{strom2018mitre}.

Taxonomies provide the domain of cybersecurity with a structured and systematic way to classify and understand threats. By having a methodological approach to organise complex information, cybersecurity researchers can improve the analysis and detection of cyber threats. The use of taxonomies as an organisational framework has become widespread in malware research for several reasons. Liles et al. \supercite{liles2015fusion} define taxonomy as a science concerned with the organisation of objects in a methodological manner. The study employs a taxonomical perspective to explore the extent to which malware can be identified and categorised as a form of weaponry. Other studies adopt a behaviour-driven classification of malware according to its type, operating platform, and method of operation \supercite{sanchez2022toward}. Similarly, another study proposed a five-dimensional taxonomy to classify the nature of cyberattacks \supercite{simmons2009avoidit}. Additional research efforts have focused on classifying malware according to its types, families, behaviours, characteristics, analysis techniques, detection methods, or associated datasets. For example, the MITRE ATT\&CK framework classifies attacks into tactics, techniques, and procedures (TTPs) based on real-world attack scenarios \supercite{strom2018mitre} to improve an organisation's defensive and offensive strategies, as well as threat actor attribution techniques. Mundie et al. \supercite{AnOntologyforMalwareAnalysis} have also developed a taxonomy of 270 malware analysis terms and subsequently integrated it into an ontology.

In the field of computer security, understanding the behaviours of malware is crucial to adequately represent knowledge and enable effective reasoning\supercite{han2021aptmalinsight}. By organising malware based on characteristics and behaviours, researchers can identify and gain new insights about various families and variants of malware by comparing them with known categories \supercite{huang2010ontology}. Organisational frameworks such as taxonomies and ontologies enable documentation of Tactics, Techniques, and Procedures (TTPs) used by malware, to aid the behavioural analysis and development of defensive capabilities to detect malware \supercite{mirza2019taxonomy, jacob2008behavioral}. It enables the prediction of adversary tactics and behaviours during attacks using recognised patterns, enhancing the abilities to mitigate and respond to future threats. In general, ontology and taxonomies have established a knowledge layer for sharing intelligence \supercite{rastogi2020malont} and best practices within the cybersecurity industry, offering a systematic method for analysing and mitigating malware, improving the overall response to cyber threats\supercite{UnifiedOntology}.
 
\subsection{Classical computation and malware}
Classical computational systems can be defined as discrete information processing models that operate through well-defined states. Each unit of information, called a ‘bit’, exists as a single state at any given moment. A binary method of computation represented as "0" and "1" forms the foundation of classical computing \supercite{deschamps2017digital}. Scientific advancements in the semiconductor industry have significantly transformed classical computational systems, enabling parallel computation through the integration of billions of transistors that help create microprocessors and other integrated circuits (IC) \supercite{sze2008semiconductor, morris1990history}. Although computational processes appear to run concurrently, computations execute sequentially at exceptional speeds, creating the illusion of simultaneous processing except in a multi-core processing environment \supercite{toong1977microprocessors, akhter2006multi} where information is computed in parallel across various processing units. This logic and reasoning form the foundation for all electronic systems, from personal computing devices to large-scale data centres \supercite{deschamps2017digital}.

A fundamental concept in classical computation originates from automata theory, which examines the nature of self-regulating mechanisms. It traces its roots back to neurophysiologists, psychologists, engineers, and mathematicians who make extensive efforts to imitate natural organisms such as a biological neuron to explore the possibilities of modelling the brain or nervous system as a kind of universal computing machine capable of responding to various external stimuli \supercite{shannon2016automata, chrisley2000artificial, mcculloch1943logical, gaines1976logic}. Automaton (singular) refers to computational models that self-regulate by existing in multiple states through the discrete sequence of inputs received from its external environment \supercite{d2012modern, burks1957logic}. 

Among these models, finite automaton (FA), also known as a finite-state machine (FSM), and Turing machines hold particular significance. FSMs are simple forms of automata consisting of a finite number of states and can be described as a five-tuple, i.e., A = (Q, I, Z, \(\partial\), W), where Q represents a set of finite states, I represents a finite collection of input symbols, Z represents a finite collection of output symbols, \(\partial\) represents transition state functions based on input, that is, (I x Q → Q). W represents the output function which maps input-state pairs to output symbols, i.e. (I x Q onto Z). A is the set of accepting states where the machine accepts an input string after processing it successfully \supercite{Aziz_Cackler_Yung, pin2010mathematical}.

\begin{figure}[H]
    \centering
    \includegraphics[width=0.6\linewidth]{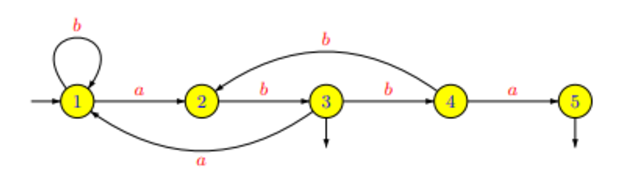} \
    \caption{Deterministic automation \supercite{pin2010mathematical}.}
    \label{fig:Deterministic-Automata}
\end{figure}

Alan Turing in 1937 extended computational capabilities beyond FSMs by introducing Turing machines to help investigate the boundaries of what can be computed \supercite{turing1936computable}. This theoretical concept employs three distinct components to execute complex algorithms: an infinite tape with separate cells to act as a memory, a control unit, and a read-write mechanism \supercite{shannon1956universal, De_Mol_Turing, Plummer_Turing}. Through Turing machines, the limits of classical computable functions can be understood, as the Turing model serves as one of the most foundational models in computational theory to advance the understanding of computable numbers and algorithmic problem solving. The principles underlying FSMs and Turing machines not only shape software and algorithm development but also influence hardware design in classical computing.

\begin{figure}[H]
    \centering
    \includegraphics[width=0.6\linewidth]{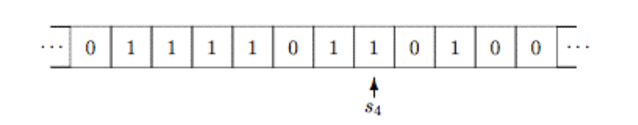} \
    \caption{Turing machine \supercite{Plummer_Turing}.}
    \label{fig:Turing-tape}
\end{figure}

Another paradigm of classical computational models that extends the principles of automata theory is cellular automata. John von Neumann in 1951, attempting to build on biological theories to conceive machines that self-reproduce by creating exact copies of itself, presented a Universal Constructor (UC) dependent on a two-dimensional cellular array consisting of 29 states per cell capable of simulating complex behaviours emerging from simple localised rule-based interactions from a neighbouring cell \supercite{chopard1998cellular, langton1984self, sarkar2000brief}.

The interconnection between theoretical computational models and the design of classical hardware components is crucial in the construction of modern computing architectures. Computational models form the central principles in the design of complex digital logic used in classical computing, including microprocessors, memory units, storage devices, control circuits, and input-output peripherals, ensuring efficient and reliable system operations. Although classical computational systems have been central to global development, underpinning the industrial revolution \supercite{mohajan2021third, fitzsimmons1994information,mowery2009plus}, they face certain limitations, particularly in offering exponential problem solving capabilities for challenges such as the factorisation of large prime numbers, which are useful for breaking current encryption algorithms \supercite{brent2000recent}. Part of this has been geared towards accelerating scientific research for more advanced computational paradigms in fields such as quantum mechanics, where atomic behaviour serves as the foundational principles for information processing models \supercite{wilde2013quantum,watrous2018theory}. By studying classical computational models and theories, researchers and engineers can enhance various aspects of existing and emerging technologies while exploring new frontiers in computer science. As technology evolves, theoretical models from classical computation can help drive innovation forward in emerging fields such as artificial intelligence, quantum technologies, and the security of complex hybrid infrastructures.  

Building on the previous discussion of classical computing models, malware often describes forms of software specifically used with the intention of exploiting or disrupting computational systems which obey classical computational theories and hardware mechanics. Although definitions and perspectives on the concepts of malware may vary, it can be concluded that malware operates within the realm of technological devices and applications, spanning tangible infrastructures as well as intangible processes that harness electromagnetic energy through scientific innovation and breakthroughs in semiconductor technology \supercite{sze2021physics}. These devices and applications include computers, software, networked systems, smartphones, and cloud-based platforms, which are built using several layers of abstraction to enable digital logic, connectivity, and functional operation. Malware often exploits vulnerabilities and existing operations in these layers to execute its malicious objectives, compromising the confidentiality, integrity, and availability of electronic systems.

\begin{figure}[H]
    \centering
    \includegraphics[width=0.9\linewidth]{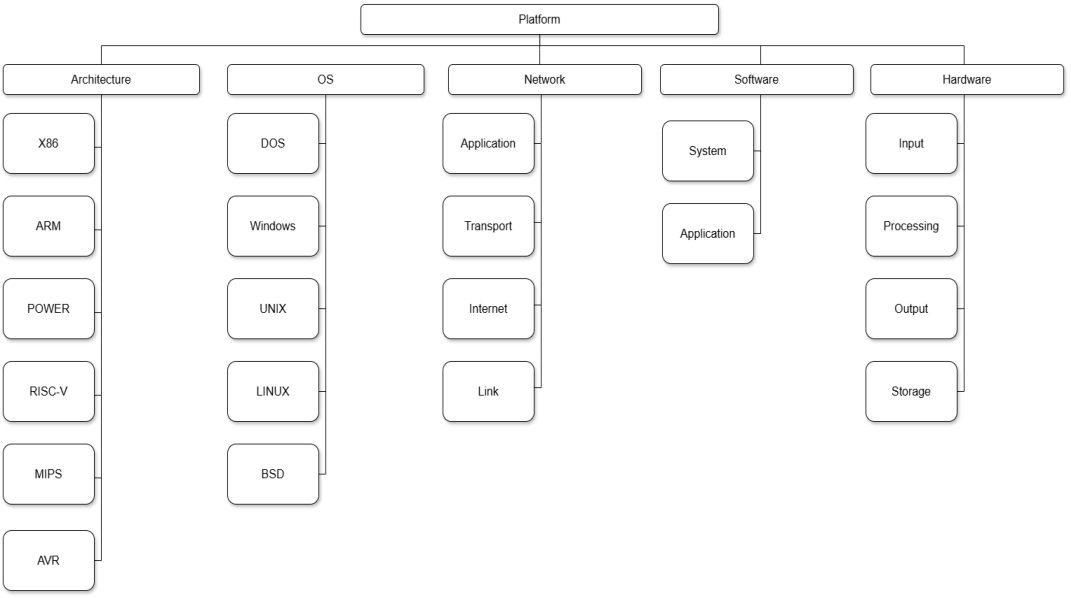} \
    \caption{Malware platforms. Adapted from \supercite{sanchez2022toward}.}
    \label{fig:classical-components}
\end{figure}

\subsection{Malware in the quantum era}
Although most of the existing literature has focused on security attacks targeting quantum architectures such as those affecting quantum algorithms, software stack, classical hardware \supercite{ghosh2023primer, baseri2024cybersecurity, johnson2019quantum, satoh2021attacking, das2023sok, saki2021qubit, lie2021hacking, gonccalves2021cyberattacks, suresh2021short}, and fault injection techniques \supercite{ash2020analysis}, relatively only a few have raised awareness of the potential risk and dangers posed by quantum malware \supercite{wu2006quantum, wang2024poster, chu2023qdoor, stolfo2021quantum}, particularly in context of the growing development and proliferation of quantum networks or the quantum Internet. This reveals a huge gap in current knowledge that warrants further investigation and contribution.

Current malware or malicious software, as discussed in the previous section, requires a specific type of computational model to execute (finite-state machines, cellular automata, and Turing machines), which helps to define applied implementations in classical semiconductor manufacturing of computing hardware. However, due to new developments and breakthroughs in material physics and semiconductor research, a range of physical platforms and quantum systems are now being explored for computational purposes, including superconducting materials \supercite{gambetta2017building, huang2020superconducting}, photonic systems\supercite{aghaee2025scaling}, and Majorana quasiparticles \supercite{microsoft2025interferometric}. Hence, the underlying hardware and physics behind computational structures are being reshaped to produce more powerful computing, sensing, and network technologies \supercite{schleich2016quantum}.

This inherently opens up a new dialogue and opportunity to examine potential security threats within such systems, particularly those involving the use of malware, allowing further exploration of the nature and existence of malware itself. To do this, it is necessary to characterise the underlying theoretical foundations, computational models, and physical principles used by these systems. Rethinking malware for quantum requires examining its potential to exploit various layers of abstraction in a quantum architecture including classical hardware components, quantum hardware, software, and quantum operations, while considering its transformation beyond conventional malicious software into entities that become integrated with or isolated within quantum operations, forming complex, interconnected, multi-domain threat structures.

\begin{figure}[H]
    \centering
    \includegraphics[width=0.6\linewidth]{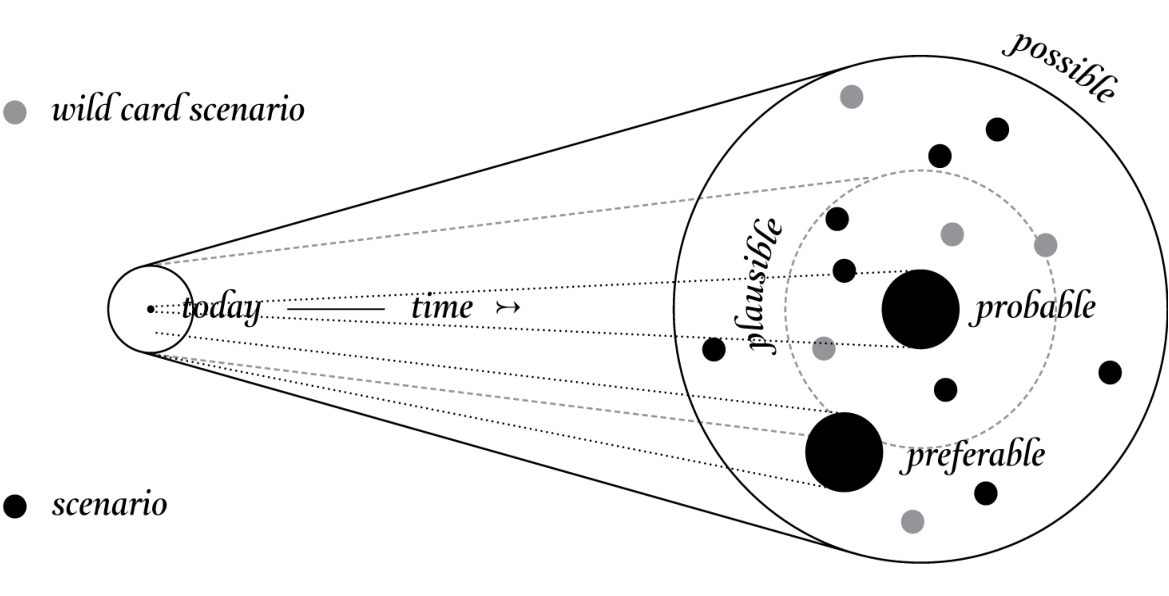} \
    \caption{Speculative future \supercite{mitrovic2016speculative}.}
    \label{fig:speculative-future}
\end{figure}

Malware will easily target or exploit vulnerabilities within classical architectures. Hence, investigating classical malware behaviours and their characteristics becomes paramount to build the necessary knowledge to guide further investigations into the unknown of possible futures. This enables the investigation of potential malicious behaviours that could be replicated against quantum architectures. Therefore, it is essential to adopt a speculative futures approach \supercite{mitrovic2016speculative} that uses innovation and invention to prototype future scenarios beyond the current concepts. This forms a critical starting point to envision how threats in quantum systems may emerge over time with the help of existing data. By anticipating these possibilities, a better understanding of the potential and limitations of what may unfold will aid in the preparation of knowledge needed to counteract threats and develop effective response strategies. In doing so, the potential to shape a more desirable and meaningful quantum future is enhanced.

\section{Research methodology}
To develop an understanding of existing malware definitions, behaviour classification, categories, and lifecycle based on ontological and taxonomical models, a systematic literature review (SLR) is conducted to support reproducibility. This methodology is shaped by careful planning and implementation based on Kitchenham's procedures \supercite{kitchenham2004procedures} for systematic reviews. However, the implementation strategy demonstrated in \supercite{jayasundara2024sok} served as a guiding reference throughout the structure and the explanation of the methodology.

\subsection{Planning the review strategy}
Creating a plan for the SLR begins with establishing the rationale for conducting the review and formulating a review protocol. This process includes several key stages designed to ensure the development of an objective and unbiased review framework.
\begin{enumerate}
  \item Establishing the research scope
  \item Formalisation of research questions
  \item Development of search terms
  \item Identification of relevant sources
  \item Study selection criteria
\end{enumerate}

\subsubsection{Establishing the research scope}
The formalisation of the scope of the research is critical in the planning phase of a systematic review of the literature (SLR) \supercite{kitchenham2004procedures}. It ensures the reliability of the research questions, the accuracy of the search terms, and the protection of the defined scope. The PICOC framework \supercite{booth2021systematic}, which is an acronym for (Population, Intervention, Comparison, Outcome, and Context), has been applied to define the scope of this SLR, as outlined in Table~\ref{tab:picoc-table}. However, given the lack of relevance for comparison in this context, conceptual mapping has been used. The scope defined subsequently informs other phases of the review process. 

\begin{table}[H]
\caption{Generation of the SLR scope guided by the PICOC framework \supercite{booth2021systematic}}
\renewcommand{\arraystretch}{1.7}
\hspace*{-2.1cm} 
\begin{tabularx}{\dimexpr\textwidth+4cm\relax}{>
{\raggedright\arraybackslash}p{0.200\textwidth} >
{\raggedright\arraybackslash}p{0.900\textwidth}}
\toprule
\setlength{\leftskip}{5em}Components & 
\setlength{\leftskip}{3em}Application to the study \\
\bottomrule
\setlength{\leftskip}{5em}Population &
\setlength{\leftskip}{3em}A focus on the classification, modelling, and representation of malware using ontologies or taxonomies \\
\setlength{\leftskip}{5em}Intervention &
\setlength{\leftskip}{3em}Application of ontological and taxonomical-based frameworks that define or categorise malware behaviours, structures, lifecycle and characteristics \\
\setlength{\leftskip}{5em}Conceptual mapping & 
\setlength{\leftskip}{3em}Exploration of how classical malware models, behaviours, and traits can be mapped into potential vulnerabilities within quantum technologies, particularly using the European Competence Framework for Quantum Technology (CFQT) as a reference for identifying architectural risk in each domain \\
\setlength{\leftskip}{5em}Outcome &
\setlength{\leftskip}{3em}A structured synthesis of existing malware classification frameworks that will inform the development of a quantum malware ontology to guide threat modelling in the quantum technology context \\
\setlength{\leftskip}{5em}Context & 
\setlength{\leftskip}{3em}The increasing convergence of quantum technologies with critical national infrastructure raises new security concerns. Understanding how classical malware could evolve or be repurposed to target quantum networks, architectures, and applications is essential for future-proofing quantum cybersecurity strategies
\end{tabularx}
\label{tab:picoc-table}
\end{table}

\subsubsection{Formalisation of research questions}
Based on the PICOC framework in Table~\ref{tab:picoc-table}, the objective of the systematic literature review (SLR) is to investigate the existing literature that describes the nature, classification, and categories of malware based on ontological and taxonomical models. Hence, the following research questions were developed.
\begin{enumerate}
    \item[\textbf{Q1:}] How can ontologies help us define and understand the fundamental nature of malware?
    \item[\textbf{Q2:}] How do taxonomies help us systematically classify malware behaviours, lifecycle, and categories?
\end{enumerate}

Q1 focuses on understanding the nature of malware through ontological knowledge concepts and relationships, while Q2 focuses on understanding how taxonomies have helped categorise malware, as well as classify their behaviours and lifecycles.

\subsubsection{Development of the search terms}
The search terms constructed to retrieve relevant studies from the appropriate sources have been derived using the population element of the PICOC framework, as shown in Table~\ref{tab:picoc-table}. Terms such as ontology and taxonomy have been chosen best to represent the formalisation of structured knowledge in malware. These terms have been selected to ensure that the SLR formally defines, categorises, and models malware using semantic or hierarchical frameworks.

Other terminology used across scholarly works to refer to malware, including its general categories such as malicious software and rogue software, were also included. These approaches helped in developing the ideal keywords and forming the search terms: "malware", "malicious software", "rogue software", "ontology", "taxonomy".

Using the advanced search query builder, the search string was constructed, incorporating fields such as ("TITLE" or "Document Title”) as a way to limit results to only articles containing the specified terms within their titles. This served as a way to ensure relevance in the retrieved data, as the review focus on central studies where malware and its classification are performed through taxonomies or ontologies, which remain vital to the research goals. This strategy helps to filter out peripheral mentions of malware, reduces noise from irrelevant articles, and enhances the particularity and quality of the results. This approach may exclude some relevant studies based on discussions in abstracts or the body of the texts, to maintain a focused and manageable review scope. To broaden the search result, a wildcard character ("*") was used in the terms malware*, malicious software*, taxonomy (taxonom*) and ontology (ontolog*) to capture all possible word variations in the search process. According to the following criteria, the search syntax applied to each of the databases is as follows:

\begin{enumerate}
  \item \textbf{Scopus:} TITLE ((malware* OR "malicious software*" OR "rogue software") AND (taxonom* OR ontolog*))

\item \textbf{IEEE Xplorer:} ("Document Title":malware* OR "Document Title":"malicious software*" OR "Document Title":"rogue software") AND ("Document Title":taxonom* OR "Document Title":ontolog*)

\item \textbf{Web of Science:} (malware* OR "malicious software*" OR "rogue software") AND (taxonom* OR ontolog*) (TITLE)
\end{enumerate}

\subsubsection{Identification of relevant sources}
The selection of the following databases: Scopus, Web of Science, and IEEE Xplore as primary data sources for this systematic literature review (SLR) stem from high indexing standards, comprehensive coverage, and relevance to the research domain. These databases are known to index peer-reviewed, high-impact research in domains such as computer science and cybersecurity, with studies on malware. Other databases, including ACM Digital Library, SpringerLink, Elsevier ScienceDirect, and Google Scholar, were considered but excluded to maintain focus and avoid duplication, as Scopus and Web of Science already index content from many of these sources. As a result, each of the chosen databases provided a robust and high-quality foundation to retrieve the relevant literature aligned with the scope defined in the PICOC framework in Table~\ref{tab:picoc-table}.

\subsubsection{Study selection criteria}
To determine how studies were selected or excluded from the SLR \supercite{kitchenham2004procedures}, the rationale for the inclusion criteria (IN) and exclusion (EX) is provided as shown in Table~\ref{tab:criteria-table}, based on the established scope presented in Table~\ref{tab:picoc-table}.

\subsection{Executing the review}
After completion of the review plan using Kitchenham's guidelines \supercite{kitchenham2004procedures}, the SLR can now be executed using the following strategies: identification and selection of appropriate studies, information extraction, and data analysis.

\subsubsection{Identification and selection of relevant studies} 
\begin{figure}[H]
    \centering
    \includegraphics[width=0.8\linewidth]{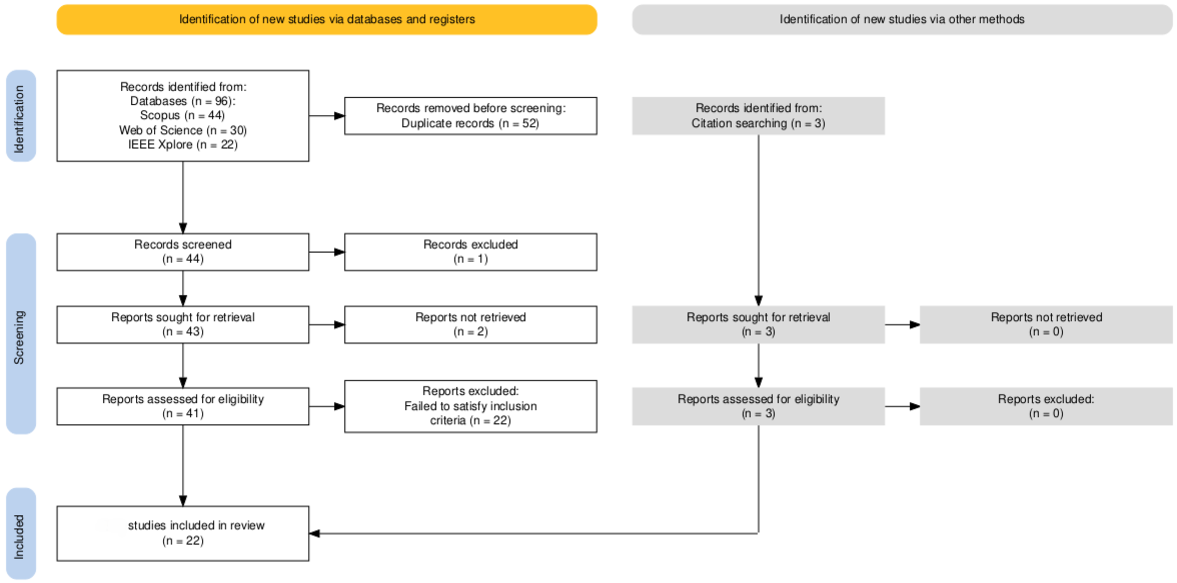} \
    \caption{PRISMA \supercite{page2021prisma} flow diagram describing the research identification, screening and inclusion process.}
    \label{fig:prismaflow}
\end{figure}

This activity was carried out from August 2024 to March 2025 according to the PRISMA framework \supercite{page2021prisma}, as shown in Figure~\ref{fig:prismaflow}. In the identification phase, a digital library search was performed that did not have any filtering of source type and date on selected academic databases (Scopus, Web of Science and IEEE Xplore), retrieving all types of scholarly records, including journal articles, book chapters, conference proceedings, white papers, editorials, and technical reports. No restrictions were placed on the type of publication to ensure comprehensive coverage of the topic.

A total of 96 articles were initially identified from all databases. After removing 52 duplicates, 44 unique articles remained for screening. Titles and abstracts were screened, but one non-English article was excluded, resulting in 43 papers sought for retrieval. Of these, 41 were successfully accessed, while two could not be recovered. After applying the inclusion and exclusion criteria by performing a thorough analysis of the text, 19 articles met the inclusion criteria and 22 were excluded.

Three articles were recovered from the snowball return search \supercite{wohlin2014guidelines} that seemed relevant to convey the main findings. Thus, a total of 22 studies were included in the final study identification and selection process, as illustrated in Figure~\ref{fig:prismaflow}.
\begin{table}[H]
\caption{Inclusion and exclusion criteria}
\renewcommand{\arraystretch}{1.7}
\hspace*{-2.1cm} 
\begin{tabularx}{\dimexpr\textwidth+4cm\relax}{>
{\raggedright\arraybackslash}p{0.100\textwidth} >
{\raggedright\arraybackslash}p{1\textwidth}}
\toprule
\setlength{\leftskip}{5em}ID & 
\setlength{\leftskip}{3em}Criteria \\
\bottomrule
\setlength{\leftskip}{5em}IN1 &
\setlength{\leftskip}{3em}The study describes the nature, behaviour, features and characteristics of malware. \\
\setlength{\leftskip}{5em}IN2 &
\setlength{\leftskip}{3em}The study focuses on highlighting various categories of malware. \\
\setlength{\leftskip}{5em}IN3 & 
\setlength{\leftskip}{3em}The study discusses malware lifecycle. \\
\setlength{\leftskip}{5em}IN4 &
\setlength{\leftskip}{3em}The study was written in English. \\
\setlength{\leftskip}{5em}IN5& 
\setlength{\leftskip}{3em}The study was peer reviewed. \\

\bottomrule
\setlength{\leftskip}{5em}EX1 & 
\setlength{\leftskip}{3em}The study focuses on malware behaviour analysis, detection techniques, and rule creation using ontology or taxonomy as a framework. This includes mobile and IoT malware. \\
\setlength{\leftskip}{5em}EX2 &
\setlength{\leftskip}{3em}The study focuses on forensic analysis with no relevance to the research questions. \\
\setlength{\leftskip}{5em}EX3 &
\setlength{\leftskip}{3em}The study focuses on developing intrusion models with no relevance to the research questions. \\
\setlength{\leftskip}{5em}EX4 & 
\setlength{\leftskip}{3em}The study lacks relevant or direct contributions to the focus of the research questions. \\

\end{tabularx}
\label{tab:criteria-table}
\end{table}

\subsubsection{Data extraction}
During the data extraction phase, a review was performed on the full texts of the articles that passed the initial screening to determine significance according to the defined inclusion criteria. Although the search strategy retrieved a wide range of papers, some focused on malware detection techniques, intrusion ontologies, mobile malware analysis (Android), digital evidence, or smart device security. Many of these were excluded, as they did not align directly with the research focus on malware ontologies and taxonomies.

Specifically, studies that focused solely on detection algorithms, digital forensics, intrusion detection systems, or analysis techniques not related to structural or conceptual modelling of malware were excluded. In contrast, papers that proposed or evaluated classification systems, taxonomies, or ontologies specific to malware behaviour, features, attack types, or analysis frameworks were retained.

For each included article, descriptive data (e.g., author, year, publication location) and analytical data relevant to the research investigation were extracted. This comprised the scope of the taxonomy or ontology, the methodology used for the construction or evaluation, the coverage of the domain, and the reuse or extension of existing conceptual frameworks.

\subsubsection{Data analysis}
To address the research questions qualitatively, a thematic analysis approach \supercite{braun2012thematic} was used. The process involved five main steps:

\begin{itemize}

\item
Familiarisation: The included studies were reviewed in NVivo, where key passages and concepts were annotated to gain initial insights into the taxonomy and ontology structures related to malware.

\item
Coding: With the research questions in focus, both deductive codes (informed by the initial review framework) and inductive codes (emerging from the data) have been applied. The codes captured aspects such as definitions, malware forms, characteristics, and malware behaviour.

\item
Defining and naming themes: The codes were organised into conceptual groups to identify recurring themes in the literature.

\item
Reviewing and refining themes: The themes were assessed and refined to ensure that the correct representation of the coded information is meaningful and correlated with the research investigation.

\item
Thematic structuring: The finalised themes became organised into higher order categories that support the structure of the findings section.
\end{itemize}

Throughout this process, codes and themes were iteratively discussed and validated to ensure consistency and alignment with research objectives.

\section{Results}
\subsection{Defining malware}
The current definition of malware is based on forms of attacks that have been repeatedly discovered and investigated over time. Many definitions lack rigour and constrain researchers to maintain a narrow perspective when defining malware properties or behaviours. However, it can be concluded that there is a focused effort on \textit{software} as the formalised domain of malware. Although the formalised and accepted definition of malware involves the use of software\supercite{maniriho2022BADTaxono} or a group of software \supercite{liles2015fusion} to breach system security or gain unauthorised access \supercite{gorment2023machine, qamar2019mobile} to silently install or execute harmful software \supercite{ismail2017general} to illicit arbitrary \supercite{Behaviouraltaxonomy} or logical behaviours on a machine or network remotely or locally to cause destruction, disruption, or theft. If malware is described as software alone, one must consider whether it has tangible aspects beyond its intangible form. As other authors have pointed out, malware can permanently destroy hardware \supercite{victor2023iot} and exploit a wide range of side channel vulnerabilities \supercite{sanchez2022toward}. There is also the intersection of malicious software triggers and their relationship with hardware trojans or in reconfigurable systems like field-programmable gate arrays (FPGAs). Malware can transcend the traditional software domain by inserting hardware Trojans that directly modify hardware behaviour. This is possible because FPGAs allow post-fabrication configuration without altering the physical structure of the silicon chip, thereby blurring categorical boundaries and calling for new definitions of malicious behaviour across the hardware–software continuum.

It remains critical to understand these nuances and relationships for research purposes, especially as these broader categories of malware can be linked or coordinated to inflict a full-spectrum attack on target infrastructures, creating far-reaching implications for the security of both present and future systems.

\subsection{Malicious behaviours}
To understand malicious behaviours broadly, several analysis and reverse engineering methods must be employed to distinguish between benign and malicious idiosyncrasies. This sets the foundation for determining what constitutes malicious or unwanted behaviour, equipping security researchers with the capability to learn about the nature of malicious attributes in a complex relationship between software and system components, including hardware, since there is no clear distinction between malicious operations and normal operations. Gr{\'e}gio et al. \supercite{gregio2012pinpointing, Behaviouralontology, Behaviouraltaxonomy, gregio2016ontology} in four separate works use the term "suspicious behaviours" to categorise these types of unwanted behaviour using ontologies and taxonomies. 

\subsubsection{Suspicious behaviours}
The first study offers a framework for understanding and detecting malware by analysing its behaviours across the entire execution context, ranging from its actions within the system to its communications over the network. It also defines the term \textit{suspicious behaviours} as a collection of operations that manipulate the condition or state of the system \supercite{gregio2012pinpointing}. 
The second and third work \supercite{Behaviouralontology, Behaviouraltaxonomy} outline the range of activities or system processes that malicious software is expected to perform on a machine, as well as its potential integration into a network. The study uses six actions to determine a suspicious event (Attack Launching, Evasion, Remote Control, Self-Defence, Stealing, and Subversion). However, it means that if a code that unintentionally exploits some specific or unknown arbitrary vulnerability (Zero Day) through arbitrary means or what is considered normal behaviour, not involving known techniques considered in the Indicators of Compromise (IoCs), the Common Attack Patterns Enumerations and Characteristics (CAPEC), or the MITRE’s Adversarial Tactics, Techniques, and Common Knowledge (ATT\&CK) \supercite{rastogi2020malont}, unfortunately, that behaviour will not be categorised as suspicious, as the criteria for Gr{\'e}gio et al.'s suspicious behaviour does not support this, making it challenging for systems to categorise except by using AI techniques, and in some cases, even such behaviours will go undetected.

\begin{figure}[H]
\includegraphics[width=15cm, height=5cm]{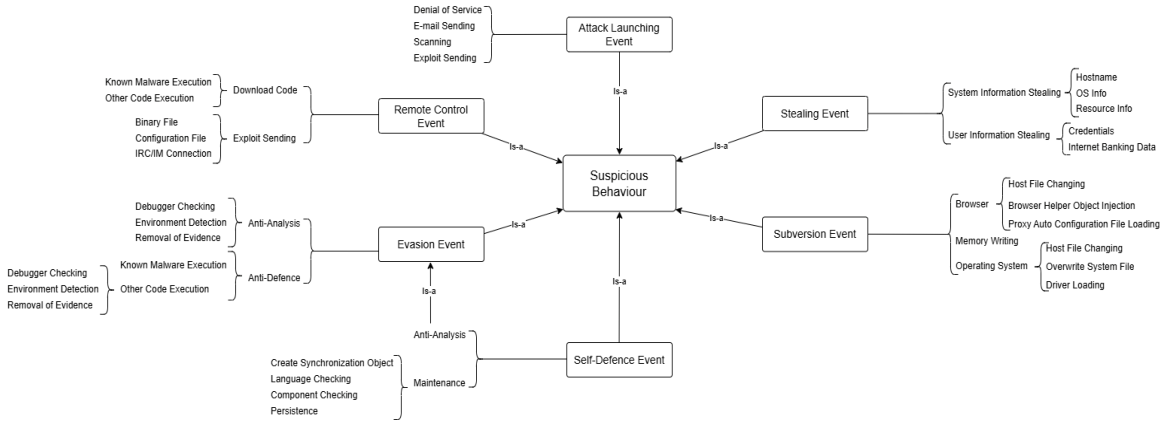}
\caption {Suspicious actions and associate behaviours. Adapted from \supercite{gregio2016ontology}.}
\centering
\end{figure}

\subsubsection{Vulnerabilities}
A more significant question can be raised: What is a vulnerability, why do systems have vulnerabilities, why do malware exploit vulnerabilities, and should vulnerabilities be considered suspicious? A system or software that has any form of vulnerability that can be exploited by some process should therefore be categorised as suspicious and a potential residence for malicious actions to occur, as such architectures naturally introduce insecurity. Considering that systems are imperfect, one must still consider the question. S{\'a}nchez-Fraga and Acosta-Bermejo \supercite{sanchez2022toward} highlight how malware exploits vulnerabilities in system design, implementation, and configuration, including weaknesses in the kernel, operating system libraries, installed programs and side channels. Kadir et al. \supercite{kadir2018understanding} suggest that malware can exploit vulnerabilities using digital communication and connectivity services, while Rastogi et al. \supercite{rastogi2020malont} point out that malware can exploit vulnerabilities present in both software and hardware. The point is that vulnerabilities can take any form and can result in intentional or unintentional weaknesses left behind by an attacker or during production of technology (hardware, software) or through ambient conditions, which might be due to corrosive wear and tear from weather conditions, i.e. affecting certain parts of the hardware system. This is why threat actors often hoard vulnerabilities and prefer to use zero-day or unknown tactics and techniques that are not yet considered suspicious by defence systems, as it increases the likelihood of evading detection. 

The capability of malware goes beyond simply compromising the integrity, availability, and confidentiality of systems and networks. Research has shown that malware is capable of disrupting nation-state activities, performing espionage activities, and sabotaging public infrastructure \supercite{Brantly01092014} \footnote{\label{conceptnote1} The works cited in [121] and [122] were not part of the systematic set, but are included as conceptual references.}. Malware, if controlled precisely, can also be used to launch weapons of mass destruction to inflict harm or cause death. This is ultimately why there is a significant arms race to accumulate vulnerabilities and exploits to leverage during conflicts \supercite{Gandhiconflict} \footref{conceptnote1} as a way to gain control over various systems. Vulnerabilities are a critical component of malware, highlighting the need for a proactive approach to address them.

\subsubsection{Unknown behaviours}
Unknown behaviours relate to uncertain transitions in computational states. This is by far the most dominant trait that malware uses to exploit security mechanisms in computational systems. By exhibiting uncertain or unrecognised behaviours, malware can effectively bypass most countermeasures. This includes exploiting new vulnerabilities or deploying evasive countermeasures \supercite{veerappan2018taxonomy}. This proves beneficial when malware tries to evade various security techniques that detect suspicious behaviours based on trained known behaviours from malware samples. However, detection systems often fail because of a lack of severe classification by inference engines. i.e. what remains a mystery or unknown cannot be inferred and judged precisely. The fourth study by Gr{\'e}gio et al. \supercite{gregio2016ontology} introduces the Malware Behaviour Ontology (MBO) to model malware behaviour knowledge and reasoning techniques to detect unknown behaviour. The framework integrates inference engines, logical rules, and risk values to detect suspicious activities during the execution of malware. Unknown behaviour is then classified into appropriate risk levels using a classification mechanism derived from these three components. These rules were created based on the behaviours of more than 10,000 malware samples analysed and captured in real world environments from the first work \supercite{gregio2012pinpointing}. The effectiveness of the rules was later evaluated using a combination of 400 benign samples and more than 2,000 malicious samples. 

A different incremental study by Huang et al. \supercite{huang2010ontology, huang2011malware, huang2012twman+} focused on creating an intelligent system to analyse malware and represent it in an ontology. The study applies several fuzzy logic frameworks \supercite{huang2011applying}, including Type-2 Fuzzy Intervals (IT2FS) \supercite{huang2014it2fs}, to model uncertainty in the malware domain characterised by high levels of imprecision and vagueness.

\subsection{Malware features and behaviour taxonomy}
Understanding the multifaceted behaviours of malware requires a multidimensional approach. Malware can have far-reaching consequences, including large-scale infrastructure damage and its use as a weapon to target critical infrastructure, such as in the case of attacks on Iran’s nuclear program \supercite{farwell2011stuxnet} \footref{conceptnote2}. In a study by Liles et al. \supercite{liles2015fusion}, the authors focus on the characteristics of weaponised malware using the deceive, disrupt, deny, degrade and destroy approach to understand the objectives of malware regarding whether it poses a weaponised threat. 

An important consideration is that malware is often thought of as having intent to be considered malware either through a designer, developer, or deployer. To simplify, just as mens rea and actus rea \supercite{sayre1932mens, smith1978actus} \footnote{\label{conceptnote2} Iran’s nuclear program [126]; see also to simplify, just as mens rea and actus rea [127], [128]; see also Singer [129] and Caltagirone et al. [130] These works were not part of the systematic set but are included as conceptual references.} is a critical component of understanding criminal acts in the physical world; malware by its nature can only be fully differentiated from poorly written or implemented code based on the intention of its design or use. If a person or AI deploys software or code without intent, which affects confidentiality, integrity, or availability, it is not malware. But if the same software or code was deployed with the intention to compromise or degrade a target system, then it is malware. The term malware, which is a portmanteu for "malicious software", amplifies the importance of malicious intent. This issue creates a problem for researchers and cybersecurity analysts, as grasping the true intention of software requires an understanding of the objective, design, or deployment of the software, which malware authors often make very difficult to accomplish. Intent can also involve different actors, meaning that the one who developed the software may be a different person from the one who deploys it, with different objectives and intentions.

Understanding the intent behind malware may require attribution in order to clarify its specific objectives. According to Singer\supercite{singer1958threat} \footref{conceptnote2}, there is a connection between the capabilities of an attacker and their intent. Within cybersecurity, this relationship is framed through the model proposed by Caltagirone et al. \supercite{caltagirone2013diamond} \footref{conceptnote2}. This diamond model of intrusion analysis explains the relationship between the attacker and the victim and the axioms relevant to this relationship. Given that a victim of malware may not be the direct intended target, the intent can become broader than a simple one-to-one relationship. For example, malware such as Stuxnet had a very specific intention and target, but the complexity of cyberspace led to a larger outbreak. Regardless of the final recipient and their relative separation from the intended target, the intent of the software itself in this case was malicious, and therefore any party who was attacked or compromised would consider it malware.

\begin{figure}[H]
  \centering
  \includegraphics[width=0.55\linewidth]{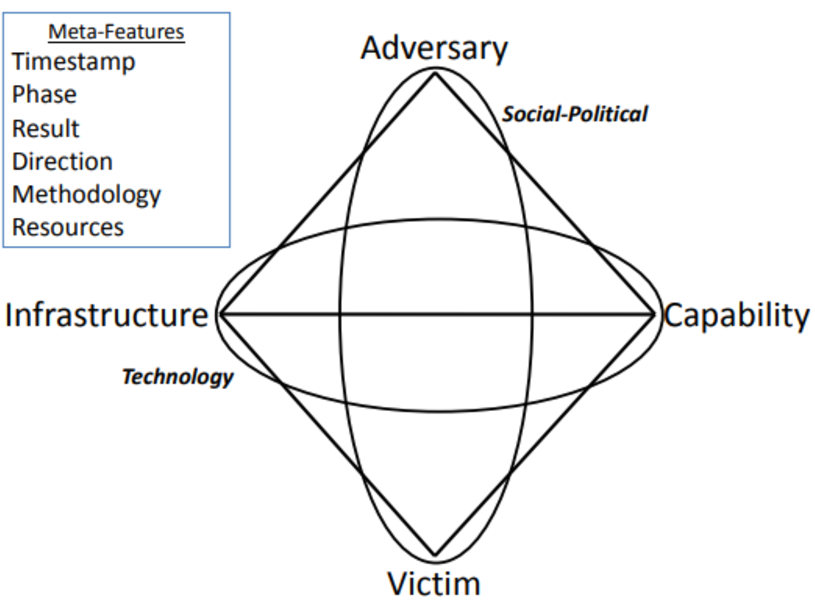}
  \caption{%
    Extended diamond model of intrusion analysis\supercite{caltagirone2013diamond}.
    \newline
    \newline
    \textit{Note:} The Social-Political link represents adversary needs and motivations 
    fulfilled by the victim, while the Technology link represents the infrastructure that enable adversary capabilities to be used in an attack. Meta-features provide contextual information about each event or activity recorded in the diamond. They do not change the four core elements, but enrich the analysis by adding dimensions such as when, how, and with what resources.
  }
  \label{fig:diamond}
\end{figure}

Malware may also exhibit benign characteristics alongside malignant behaviours, depending on its overarching goals, as a strategy to conceal its true intent from both defence mechanisms and cybersecurity analysts. Table~\ref{tab:Mbehaviourtable} presents a taxonomy of malware behaviours and features compiled from multiple sources in the literature. Seven categories of behaviours are identified: (1) \textit{Weaponization}, which refers to the consequences of malware deployment; (2) \textit{Evasion}, describing techniques used to bypass security mechanisms; (3) \textit{Code organisation}, outlining how malware code is structured and assembled; (4) \textit{Advanced persistence}, detailing the methods employed to maintain long-term and unrestricted access; (5) \textit{Operations}, referring to normal system functions leveraged for malicious activity; (6) \textit{Deception}, encompassing methods used to mislead system processes; and (7) \textit{Dynamic and memory features}, which describe how malware interacts with and manipulates memory.

\begin{table}[H]
\vspace{-1cm}
\caption{Taxonomy of malware behaviour and features}
\renewcommand{\arraystretch}{1.7}
\hspace*{-2.1cm} 
\begin{tabularx}{\dimexpr\textwidth+4cm\relax}{>{\raggedright\arraybackslash}p{0.250\textwidth} >{\raggedright\arraybackslash}p{0.250\textwidth} >{\raggedright\arraybackslash}p{0.300\textwidth} >{\raggedright\arraybackslash}p{0.350\textwidth}}
\toprule
Attributes & Platform & Behaviour & Description \\
\bottomrule

& & Deceive & Mislead by false representation of data \\
& & Disrupt & Create disorder in regular computer functionality \\
\textbf{Weaponisation} \supercite{liles2015fusion} & Generic & Deny & Withhold or block information processing services \\
& & Degrade & Lowering quality or functionality of networks and systems  \\
& & Destroy & Render ineffective, useless, cause injury/death or completely defeat to become unusable \\
\midrule

& & 
\begin{enumerate}
\vspace{-1.9em}
\setlength{\leftskip}{-2.5em}
        \item [{}] (Sandbox evasion)
         \item [{}] \textit{Sleep}
\renewcommand\labelitemi{\tiny$\bullet$}
\begin{itemize}
\setlength{\leftskip}{-3em}
   \vspace{-0.1cm}\item Extended sleep
    \vspace{-0.1cm}\item Stalling code
\end{itemize} 

\item[{}] \textit{User behaviour spoofing}
\begin{itemize}
\setlength{\leftskip}{-3em}
   \vspace{-0.1cm}\item User interaction
    \vspace{-0.1cm}\item Abnormal macro
\end{itemize}

\item[{}] \textit{Event based}
\begin{itemize}
\setlength{\leftskip}{-3em}
   \vspace{-0.1cm}\item Sandbox event
\end{itemize}

\item[{}] \textit{Configuration based spoofs}
\begin{itemize}
\setlength{\leftskip}{-3em}
   \vspace{-0.1cm}\item Leveraging specific information 
    \vspace{-0.1cm}\item Environment configuration
\end{itemize}
\end{enumerate}
& 
\begin{enumerate} 
\vspace{-1.9em}
\setlength{\leftskip}{-2.5em}
        \item [{}] (Avoid detection in isolated setups)
         \item [{}] \textit{Wait out the anti-malware analysis}
\renewcommand\labelitemi{\tiny$\bullet$}
\begin{itemize}
\setlength{\leftskip}{-3em}
   \vspace{-0.1cm}\item DelayInterval INFINITE
    \vspace{-0.1cm}\item Sleep () function
\end{itemize} 

\item[{}] \textit{Indicative of user activity}
\begin{itemize}
\setlength{\leftskip}{-3em}
   \vspace{-0.1cm}\item Detect number of mouse clicks
    \vspace{-0.1cm}\item Application.RecentFiles.Count
\end{itemize}

\item[{}] \textit{User action requirement}
\begin{itemize}
\setlength{\leftskip}{-3em}
   \vspace{-0.1cm}\item Date/time checks
\end{itemize}

\item[{}] \textit{Indicative of a VM based sandbox}
\begin{itemize}
\setlength{\leftskip}{-3em}
   \vspace{-0.1cm}\item Existence of certain artefacts, i.e, files, processes, drivers
    \vspace{-0.1cm}\item Executes only on intended target system configuration
\end{itemize}
\end{enumerate}
\\

\vspace{-8em}\textbf{Evasion} \supercite{veerappan2018taxonomy} & \vspace{-8em} Generic & 
\begin{enumerate}
\vspace{-2.5em }
\setlength{\leftskip}{-2.5em}
    \vspace{-1em}\item [{}] (Anti-disassembly evasion)
\renewcommand\labelitemi{\tiny$\bullet$}
\begin{itemize}
\setlength{\leftskip}{-3em}
   \vspace{-0.1cm}\item Flow - oriented disassembler techniques
    \vspace{-0.1cm}\item Code obfuscation
\end{itemize} 
\end{enumerate} 
& 
\begin{enumerate} 
\vspace{-2.5em} 
\setlength{\leftskip}{-2.5em}
    \vspace{-1em}\item [{}] (Delay or prevent code analysis)
\renewcommand\labelitemi{\tiny$\bullet$}
\begin{itemize}
\setlength{\leftskip}{-3em}
   \vspace{-0.1cm}\item Double jump, garbage bytes, return pointer manipulation
    \vspace{-0.1cm}\item ROT13, dead code insertion packers, base64 encoding
\end{itemize} 
\end{enumerate} 
\\

& & 
\begin{enumerate}
\vspace{-2.5em} 
\setlength{\leftskip}{-2.5em}
    \vspace{-1em}\item [{}] (Anti-Debugging Evasion)
\renewcommand\labelitemi{\tiny$\bullet$}
\begin{itemize}
\setlength{\leftskip}{-3em}
   \vspace{-0.1cm}\item Anti-debugging
    \vspace{-0.1cm}\item Code inspection \& patching
\end{itemize} 
\vspace{-1.5em} 
\end{enumerate} 
& 
\begin{enumerate} 
\vspace{-2.5em} 
\setlength{\leftskip}{-2.5em}
    \vspace{-1em}\item [{}] (Detect \& disrupt debugger)
\renewcommand\labelitemi{\tiny$\bullet$}
\begin{itemize}
\setlength{\leftskip}{-3em}
   \vspace{-0.1cm}\item IsDebuggerPresent (API)
    \vspace{-0.1cm}\item Exploits (SEH), (VEH) or (UEF)
\end{itemize} 
\vspace{-1.5em} 
\end{enumerate} 
\\
\midrule

& & Basic & No techniques to hide or alter its behaviour \\
& & Metamorphic & Completely rewrites itself each execution \\
\textbf{Code organisation} \supercite{chowdhury2022capturing} & Generic & Polymorphic & Same function, different appearance each execution \\
& & Packing & Compress or encrypt original code \\
& & Code obfuscation & Conceal malware activities \\

\end{tabularx}
\label{tab:Mbehaviourtable}
\end{table}

\begin{table}[H]
\vspace{-1cm}
\caption*{{Table 3: }(\textit{continued)}}
\renewcommand{\arraystretch}{1.7}
\hspace*{-2.1cm} 
\begin{tabularx}{\dimexpr\textwidth+4cm\relax}{>{\raggedright\arraybackslash}p{0.250\textwidth} >{\raggedright\arraybackslash}p{0.250\textwidth} >{\raggedright\arraybackslash}p{0.300\textwidth} >{\raggedright\arraybackslash}p{0.350\textwidth}}
\toprule
Attributes & Platform & Behaviour & Description \\
\bottomrule

\vspace{8em}\textbf{Advanced Persistence} \supercite{han2021aptmalinsight} & \vspace{8em} Generic & 
\begin{enumerate}
\vspace{-1.7em}
\setlength{\leftskip}{-2.5em}
         \item [{}] \textit{Remote control}
\renewcommand\labelitemi{\tiny$\bullet$}
\begin{itemize}
\setlength{\leftskip}{-3em}
    \vspace{-0.1cm}\item Connect C\&C server
    \vspace{-0.1cm}\item Get command
    \vspace{-0.1cm}\item Download code
\end{itemize} 

\item[{}] \textit{Subversion}
\begin{itemize}
\setlength{\leftskip}{-3em}
    \vspace{-0.1cm}\item Memory writing
    \vspace{-0.1cm}\item File modification
    \vspace{-0.1cm}\item Registry modification
\end{itemize}

\item[{}] \textit{Stealing}
\begin{itemize}
\setlength{\leftskip}{-3em}
   \vspace{-0.1cm}\item System information 
    \vspace{-0.1cm}\item User information
\end{itemize}

\item[{}] \textit{Self defence}
\begin{itemize}
\setlength{\leftskip}{-3em}
    \vspace{-0.1cm}\item Environment detection
    \vspace{-0.1cm}\item Anti-analysis
    \vspace{-0.1cm}\item Self-deletion
\end{itemize}
\vspace{-1.5em}
\end{enumerate}
& 
\begin{enumerate} 
\vspace{-1.7em}
\setlength{\leftskip}{-2.5em}
         \item [{}] \textit{Compromised system control}
\renewcommand\labelitemi{\tiny$\bullet$}
\begin{itemize}
\setlength{\leftskip}{-3em}
   \vspace{-0.1cm}\item Communication channel
    \vspace{-0.1cm}\item Retrieving data
     \vspace{-0.1cm}\item Downloading data
\end{itemize} 

\item[{}] \textit{Undermine or weaken policy}
\begin{itemize}
\setlength{\leftskip}{-3em}
   \vspace{-0.1cm}\item Inject code into RAM memory
    \vspace{-0.1cm}\item Modify system files
     \vspace{-0.1cm}\item Modify system policies
\end{itemize}

\item[{}] \textit{Pilfer data}
\begin{itemize}
\setlength{\leftskip}{-3em}
   \vspace{-0.1cm}\item Operating system information
    \vspace{-0.1cm}\item Password, PII
\end{itemize}

\item[{}] \textit{Prevent detection and termination}
\begin{itemize}
\setlength{\leftskip}{-3em}
   \vspace{-0.1cm}\item Detect sandboxes or VMs
    \vspace{-0.1cm}\item Obstruct code analysis
     \vspace{-0.1cm}\item Erase evidence
\end{itemize}
\vspace{-1.5em}
\end{enumerate}
\\
\midrule

\vspace{15em}\textbf{Operations} \supercite{sanchez2022toward} & \vspace{15em} Generic & 
\begin{enumerate}
\vspace{-1.7em}
\setlength{\leftskip}{-2.5em}
         \item [{}] \textit{Configuration manipulation}
\renewcommand\labelitemi{\tiny$\bullet$}
\begin{itemize}
\setlength{\leftskip}{-3em}
    \vspace{-0.1cm}\item User application
    \vspace{-0.1cm}\item Operating system
    \vspace{-0.1cm}\item Library/module
    \vspace{-0.1cm}\item Firmware  
\end{itemize} 

\item[{}] \textit{Information manipulation}
\begin{itemize}
\setlength{\leftskip}{-3em}
    \vspace{-0.1cm}\item Distortion
    \vspace{-0.1cm}\item Disruption
    \vspace{-0.1cm}\item Destruction
    \vspace{-0.1cm}\item Disclosure
    \vspace{-0.1cm}\item Discover
\end{itemize}

\item[{}] \textit{Misuse of resources}
\begin{itemize}
\setlength{\leftskip}{-3em}
    \vspace{-0.1cm}\item Processor 
    \vspace{-0.1cm}\item Storage
    \vspace{-0.1cm}\item Network
\end{itemize}

\item[{}] \textit{Network communication}
\begin{itemize}
\setlength{\leftskip}{-3em}
    \vspace{-0.1cm}\item HTTP request
    \vspace{-0.1cm}\item DNS resolution
    \vspace{-0.1cm}\item IP traffic
\end{itemize}

\item[{}] \textit{File system manipulation}
\begin{itemize}
\setlength{\leftskip}{-3em}
    \vspace{-0.1cm}\item Open
    \vspace{-0.1cm}\item Write
    \vspace{-0.1cm}\item Delete
\end{itemize}

\item[{}] \textit{Process manipulation}
\begin{itemize}
\setlength{\leftskip}{-3em}
    \vspace{-0.1cm}\item Creation
    \vspace{-0.1cm}\item Kill
    \vspace{-0.1cm}\item Injection
\end{itemize}
\end{enumerate}
&  
\begin{enumerate} 
\vspace{-1.7em}
\setlength{\leftskip}{-2.5em}
         \item [{}] \textit{Settings modification}
\renewcommand\labelitemi{\tiny$\bullet$}
\begin{itemize}
\setlength{\leftskip}{-3em}
   \vspace{-0.1cm}\item Adjust user binary preference
    \vspace{-0.1cm}\item Modify system-level settings
     \vspace{-0.1cm}\item Modify software variables
      \vspace{-0.1cm}\item Modify low-level software
\end{itemize} 

\item[{}] \textit{Data modification}
\begin{itemize}
\setlength{\leftskip}{-3em}
   \vspace{-0.1cm}\item Alter information
    \vspace{-0.1cm}\item Interrupt information flow
     \vspace{-0.1cm}\item Damage information
      \vspace{-0.1cm}\item Leak information
       \vspace{-0.1cm}\item Uncover information
\end{itemize}

\item[{}] \textit{Unauthorised inefficient use}
\begin{itemize}
\setlength{\leftskip}{-3em}
   \vspace{-0.1cm}\item Consumes CPU cycles/power
    \vspace{-0.1cm}\item Excessive use of storage space
     \vspace{-0.1cm}\item Misuse of network bandwidth
\end{itemize}

\item[{}] \textit{Data connectivity}
\begin{itemize}
\setlength{\leftskip}{-3em}
   \vspace{-0.1cm}\item Camouflage within web traffic
    \vspace{-0.1cm}\item Traffic redirection
     \vspace{-0.1cm}\item Service communication
\end{itemize}

\item[{}] \textit{File modification}
\begin{itemize}
\setlength{\leftskip}{-3em}
   \vspace{-0.1cm}\item Access files
    \vspace{-0.1cm}\item Add malicious data to files
     \vspace{-0.1cm}\item Erase files
\end{itemize}

\item[{}] \textit{Prevent detection and termination}
\begin{itemize}
\setlength{\leftskip}{-3em}
   \vspace{-0.1cm}\item Initiate new process
    \vspace{-0.1cm}\item Terminate existing process
     \vspace{-0.1cm}\item Insert code into a process
\end{itemize}
\end{enumerate}
\\
\end{tabularx}
\end{table}

\begin{table}[H]
\vspace{-1cm}
\caption*{{Table 3: }(\textit{continued)}}
\renewcommand{\arraystretch}{1.7}
\hspace*{-2.1cm} 
\begin{tabularx}{\dimexpr\textwidth+4cm\relax}{>{\raggedright\arraybackslash}p{0.250\textwidth} >{\raggedright\arraybackslash}p{0.250\textwidth} >{\raggedright\arraybackslash}p{0.300\textwidth} >{\raggedright\arraybackslash}p{0.350\textwidth}}
\toprule
Attributes & Platform & Behaviour & Description \\
\bottomrule

& & Camouflage & Disguise malware presence \\
& & Encryption & Convert plain text to ciphertext \\
& & Oligomorphic & Changes its code to a limited extent \\
& & Polymorphic & Same function, different appearance each execution \\
\textbf{Deception} \supercite{gorment2023machine} & Generic & Metamorphic & Completely rewrites itself each execution \\
&  & 
\begin{enumerate}
\vspace{-1.7em}
\setlength{\leftskip}{-2.5em}
         \item [{}] \textit{Obfuscation}
\renewcommand\labelitemi{\tiny$\bullet$}
\begin{itemize}
\setlength{\leftskip}{-3em}
    \vspace{-0.1cm}\item Dead code insertion
    \vspace{-0.1cm}\item Instruction replacement
    \vspace{-0.1cm}\item Register reassignment
    \vspace{-0.1cm}\item Subroutine reordering
    \vspace{-0.1cm}\item Code transposition
    \vspace{-0.1cm}\item Code integration
\end{itemize} 
\vspace{-1.5em} 
\end{enumerate}
& 
\begin{enumerate} 
\vspace{-1.7em}
\setlength{\leftskip}{-2.5em}
         \item [{}] \textit{Hide malware's nature \& function}
\renewcommand\labelitemi{\tiny$\bullet$}
\begin{itemize}
\setlength{\leftskip}{-3em}
   \vspace{-0.1cm}\item Adds non-functional code 
    \vspace{-0.1cm}\item Swaps equivalent instruction
     \vspace{-0.1cm}\item Changing code registers
      \vspace{-0.1cm}\item Shuffles order of subroutines
       \vspace{-0.1cm}\item Rearrange order of code blocks
        \vspace{-0.1cm}\item Merges with legitimate code
\end{itemize}
\vspace{-1.5em} 
\end{enumerate}
\\
\midrule

\vspace{8em}\textbf{Dynamic and memory features} \supercite{maniriho2022BADTaxono} & \vspace{8em} Generic & 
\begin{enumerate}
\vspace{-1.7em}
\setlength{\leftskip}{-2.5em}
         \item [{}] \textit{Systematic behaviours}
\renewcommand\labelitemi{\tiny$\bullet$}
\begin{itemize}
\setlength{\leftskip}{-3em}
    \vspace{-0.1cm}\item API sequence calls
    \vspace{-0.1cm}\item API call frequencies
    \vspace{-0.1cm}\item API return values
    \vspace{-0.1cm}\item Registry operations
    \vspace{-0.1cm}\item Loaded DLLs
    \vspace{-0.1cm}\item DLL function calls
    \vspace{-0.1cm}\item File system changes (operation on file)
    \vspace{-0.1cm}\item Mutexes’ operations-based features
    \vspace{-0.1cm}\item Values of register
    \vspace{-0.1cm}\item File changes in suspicious locations
    \vspace{-0.1cm}\item Suspicious DLL location
    \vspace{-0.1cm}\item Process
    \vspace{-0.1cm}\item Threads
\end{itemize} 

\item[{}] \textit{Network behaviours}
\begin{itemize}
\setlength{\leftskip}{-3em}
    \vspace{-0.1cm}\item Network operations
    \vspace{-0.1cm}\item Web browsing history
\end{itemize}
\vspace{-1.5em}
\end{enumerate}
& 
\begin{enumerate} 
\vspace{-1.7em}
\setlength{\leftskip}{-2.5em}
         \item [{}] \textit{Organised system behaviours}
\renewcommand\labelitemi{\tiny$\bullet$}
\begin{itemize}
\setlength{\leftskip}{-3em}
   \vspace{-0.1cm}\item Suspicious API requests
   \vspace{-0.1cm}\item Recurrence of API requests
   \vspace{-0.1cm}\item Outcome of each API calls
   \vspace{-0.1cm}\item Abuse of system configuration
   \vspace{-0.1cm}\item Map the DLL into memory
   \vspace{-0.1cm}\item DLL functions imported
   \vspace{-0.1cm}\item File manipulation – copy, delete, rename, rewrite
   \vspace{-0.1cm}\item Executing Mutual exclusion objects
   \vspace{-0.1cm}\item Data values from registers
   \vspace{-0.1cm}\item File modification in sensitive locations
   \vspace{-0.1cm}\item Unrecognised DLL path
   \vspace{-0.1cm}\item Insertion of corrupt process
   \vspace{-0.1cm}\item A unit of malicious instruction

\end{itemize} 

\item[{}] \textit{Host communication}
\begin{itemize}
\setlength{\leftskip}{-3em}
   \vspace{-0.1cm}\item open ports, sockets, IP address
    \vspace{-0.1cm}\item Connection to malicious Urls
\end{itemize}
\end{enumerate}
\\
\end{tabularx}
\end{table}

\subsection{Malware forms, families and variants}
There are several well-documented types of malware in the literature that help categorise various forms of malware according to their expected behaviour, functionalities, or intentions. However, creating a straightforward taxonomy to identify and group malware remains challenging because malware classes often overlap in their behaviours, while others may be variations of certain malicious activities and classes \supercite{Behaviouraltaxonomy}. According to an extensive literature review conducted by Gr{\'e}gio et al. \supercite{Behaviouraltaxonomy}, malware can be categorised into five distinct categories: (viruses/worms/trojans, botnets, spyware, rootkits and surreptitious software). Another study aimed at developing a knowledge base through an ontology of malware families and individual classes through a literature review \supercite{ding2019ontology} provides a reference that classifies malware into seven categories: (viruses, trojans, backdoors, rogue software, time bomb, and logic bomb). However, the study acknowledges the classification of malware into five categories: (viruses, trojans, worms, backdoors, and rootkits). Although reference from Gr{\'e}gio et al. \supercite{Behaviouraltaxonomy} indicates that viruses, worms, and trojans are infection programs with overlapping behaviours, dividing them into two distinct classes: (logic bombs/trojans) and (viruses/worms). Gr{\'e}gio et al argues that worms can be considered a form of virus when viewed through an algorithmic lens. Due to the significant number of malware samples produced each day by malware authors, the study by Yuxin et al. \supercite{ding2019ontology} suggests that malware can be divided into subcategories, families, and individuals (instances or variants). It defines malware individuals as having specific characteristics or code features related to a malware family, while malware families consist of groups of malware that exhibit similar behaviours or share common features or code in their samples.

Recently, more sophisticated forms of malware have been designed to exhibit novel attack methods targeting conventional and modern architectures, such as IoT \supercite{victor2023iot, 10yearsofIoTmalware, vignau2021evolution}. These are represented in the forms of (Adware, ransomware, information stealers, bots/botnets, fileless malware, hybrid malware, keyloggers, droppers, downloaders and cryptominer) \supercite{maniriho2022BADTaxono, gorment2023machine, chowdhury2022capturing}.

With significant technological advancements leading to the development of smartphones, particularly the widespread adoption of the Android operating system, which has more than 6 billion users worldwide, mobile malware has become a persistent threat. This is due to the critical and unique information stored on these devices. A study by Ismail et al. \supercite{ismail2017general} categorises Android malware into six distinct classes: (Adware, Bot, Rootkit, Spyware, Trojan, Worm) based on previous research in the literature, while another study by Kadir et al. \supercite{kadir2018understanding} expands the categories of financial malware in Android smartphones into five groups consisting of Adware, Banking Malware, Ransomware, SMS Malware, and Scareware. In more recent research on mobile malware, the study by Qamar et al. \supercite{qamar2019mobile} classifies malware into 10 types: (virus, worm, trojan, rootkit, botnet, adware, spyware, ransomware, backdoors, and keyloggers). With the adaptable nature of malware, new forms and techniques are likely to continue to emerge to meet the various expectations of malicious actors.

Hence, ontologies provide a stronger basis than taxonomies for classifying and identifying malware, as they can account for both observed and unobserved behaviours or similarities across forms, families, and samples. Also, because malware behaviours often inherit or reuse features from one another, detecting intent becomes more challenging, further highlighting the value of ontological approaches.

\begin{table}[H]
\vspace{-1cm}
\caption{Categories and malware families}
\renewcommand{\arraystretch}{1.7}
\hspace*{-2.1cm} 
\begin{tabularx}{\dimexpr\textwidth+4cm\relax}{>
{\raggedright\arraybackslash}p{0.250\textwidth} >
{\raggedright\arraybackslash}p{0.250\textwidth} >
{\raggedright\arraybackslash}p{0.650\textwidth}}
\toprule
Categories & Examples of Known Families & Description \\
\bottomrule
Virus & Amnesia \supercite{victor2023iot}, Yale, Stoned, Melissa \supercite{gorment2023machine} & Self-replicating malicious code that propagates by attaching itself to other legitimate programs or files after execution \supercite{Behaviouraltaxonomy}\supercite{maniriho2022BADTaxono}\supercite{gorment2023machine} \\\hline
Trojan & Emotet \supercite{victor2023iot} ZeuS \supercite{gorment2023machine} & Designed to appear as a legitimate program to trick users into executing the program \supercite{Behaviouraltaxonomy}\supercite{maniriho2022BADTaxono}\supercite{gorment2023machine} \\\hline
Worms & Morris, Stuxnet \supercite{gorment2023machine} & Designed to spread across devices over a network without any form of human operation during program execution \supercite{Behaviouraltaxonomy}\supercite{maniriho2022BADTaxono}\supercite{gorment2023machine} \\\hline
Backdoor (Remote access trojan) & Brickerbot \supercite{victor2023iot} Titanium APT \supercite{gorment2023machine} & Used to provide remote network access to attackers to assist in subsequent system compromise and infection \supercite{maniriho2022BADTaxono}\supercite{gorment2023machine} \\\hline
Spyware & VPNFilter \supercite{victor2023iot} & Specialises in spying and monitoring behavioural activity of users without their knowledge or consent to collect keystrokes, screenshots, passwords, logs, etc. \supercite{Behaviouraltaxonomy}\supercite{maniriho2022BADTaxono}\supercite{gorment2023machine} \\\hline
Botnet & Mirai, \supercite{gorment2023machine} & Used for infecting multiple devices to create a swarm of bots, giving remote access and control to an attacker 
(bot master). It is primarily used in performing denial of service (DOS/DDOS) services \supercite{Behaviouraltaxonomy}\supercite{maniriho2022BADTaxono}\supercite{gorment2023machine} \\\hline
Rootkit & TDL3 \supercite{gorment2023machine} & A collection of malicious programs designed to perform a range of functionalities, stay hidden and use administrative privileges to gain access and control systems. \supercite{Behaviouraltaxonomy}\supercite{maniriho2022BADTaxono}\supercite{gorment2023machine} \\\hline
Surreptitious software & & Exploits obfuscation techniques to prevent users from reverse engineering and figuring out how to defeat its
malicious activities \supercite{Behaviouraltaxonomy} \\\hline
Rogue software & & Targets unsuspecting users by pretending to be a legitimate security software to trick users into downloading and installing it \supercite{ding2019ontology} \supercite{chowdhury2022capturing} \\\hline
Logic bomb & & Executes and performs malicious activities based on a specific set of logic triggered by user or system activities \supercite{ding2019ontology} \\\hline
Time bomb & & Operates by executing malicious activities according to predefined schedules \supercite{ding2019ontology} \\\hline
Adware & Shlayer \supercite{gorment2023machine} & Serves pop-up adverts on browsers and systems to generate cash or steal user information when clicked on \supercite{maniriho2022BADTaxono}\supercite{gorment2023machine}\\\hline
Information stealer & & Designed for stealing confidential information such as keystrokes and banking credentials \supercite{maniriho2022BADTaxono} \\\hline
Ransomware & Hades \supercite{victor2023iot} Wannacry \supercite{gorment2023machine} & Specialises in encrypting files, blocking access until a ransom, often paid in bitcoin, is received \supercite{maniriho2022BADTaxono}\supercite{gorment2023machine} \\\hline
Fileless malware & & Volatile memory-based malware that perform its operations without writing to disk (Physical drive) to stay undetectable \supercite{maniriho2022BADTaxono} \\\hline
\end{tabularx}
\end{table}

\begin{table}[H]
\vspace{-1cm}
\caption*{{Table 4: }(\textit{continued)}}
\renewcommand{\arraystretch}{1.7}
\hspace*{-2.1cm} 
\begin{tabularx}{\dimexpr\textwidth+4cm\relax}{>
{\raggedright\arraybackslash}p{0.250\textwidth} >
{\raggedright\arraybackslash}p{0.250\textwidth} >
{\raggedright\arraybackslash}p{0.650\textwidth}}
\toprule
Categories & Families & Descriptions \\
\bottomrule
Hybrid malware & & Incorporates various malware functionalities into one to enhance capabilities and attack \supercite{maniriho2022BADTaxono} \\\hline
keyloggers & & Captures keystrokes of users while typing to steal sensitive and confidential information such as passwords \supercite{gorment2023machine} \\\hline
Dropper & & Trojan designed to deliver other types of malware to a system \supercite{chowdhury2022capturing} \\\hline
Downloader & & Used to download malware used for multiple stages of infection from the internet \supercite{chowdhury2022capturing} \\\hline
Cryptominer & & Exploits victim's computational resources to mine cryptocurrency \supercite{chowdhury2022capturing} \\\hline
Banking malware & Citmo, Bankbot \supercite{kadir2018understanding} & Designed to mimic a legitimate banking app to lure and exploit victims \supercite{kadir2018understanding} \\\hline
SMS malware & Gazon, Uxipp \supercite{kadir2018understanding} & Targets mobile phones by using sms services as a means of delivering malicious payload \supercite{kadir2018understanding} \\\hline
Scareware & Avpass, FakeAV \supercite{kadir2018understanding} & Deceive users into thinking their computer is infected, luring scared users into buy fake solutions that install actual malware  \supercite{kadir2018understanding} \\\hline
\end{tabularx}
\end{table}

\subsection{Malware attack lifecycle}
The deployment of malware often requires a specific set of tactics and strategies to be successful, especially when targeting highly secure systems or infrastructures.  A research contribution by Gorment et al. \supercite{gorment2023machine} demonstrates how malware attacks usually occur in phases. Each attack phase can vary or be unique to the type of malware and family. Although each attack phase is not strictly sequential, the lifecycle of most malware attacks typically adheres to a recognisable pattern, as the study shows. A malware attack often begins by identifying the target infrastructure to find the weakest link and potential vulnerabilities, and assessing defences to choose an effective attack tactic and strategy. The next stage involves exploiting the vulnerabilities discovered in the systems and bypassing security mechanisms to gain access to user accounts and systems on the network. After a successful break-in and attackers have gained access to the network, the next stage involves setting up remote access to establish and ensure two-way communication between the attacker device and the compromised network. The next stage is an infection stage that occurs after remote access is secured. This phase involves spreading the attack or infecting other systems by compromising additional user accounts, high-privileged accounts, and network segments. The final phase of a malware attack consists of achieving the intended objectives, which may include stealing, encrypting, destroying, or corrupting data or the system. 

\begin{figure}[ht!]
    \centering
    \includegraphics[width=0.8\linewidth]{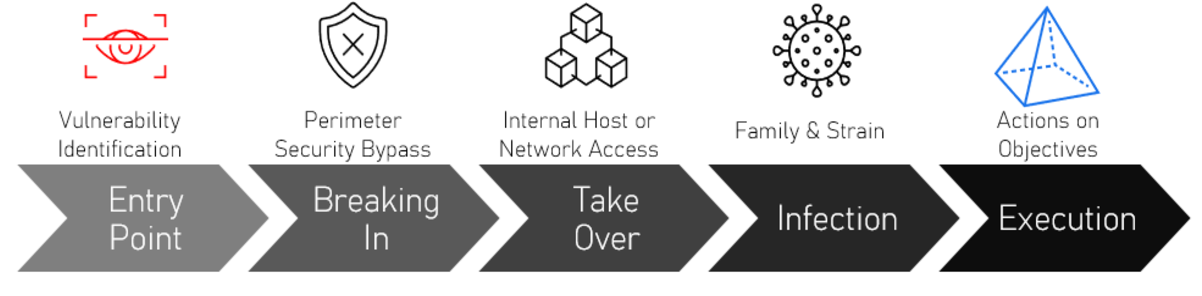}
    \caption{Malware attack chain and lifecycle. Adapted from \supercite{gorment2023machine}.}
    \label{fig:enter-label}
\end{figure}

\subsection{Malware ATT\&CK matrices}
Malware's multidimensional and complex approach in attacking target systems and infrastructures means that simple attack lifecycles such as the one described earlier or similar to the Lockheed Martin cyber kill chain \supercite{naik2022comparing} \footnote{\label{conceptnote3} Lockheed Martin cyber kill chain [134]; see also adversary behaviour and tactics [135]; see also attack attribution [136]; see also required to achieve operational goal [137]; see also victim’s internal network [138]. These works were not part of the systematic set but are included as conceptual references.} are insufficient to fully grasp the depth of attack techniques and strategies frequently employed by malicious threat actors, particularly Advanced Persistent Threats (APT) actors. These types of malware typically involve sophisticated attack campaigns orchestrated by groups or individuals skilled in programming malicious code, evading defences, and assessing vulnerabilities deep within the system or network. This often results in the creation of new malware forms, including families and variants, as malware development requires creativity and a deep understanding of computer behaviours. 

Discovery and understanding of adversary behaviours, motives, and objectives are essential to formalise malware knowledge, which, in turn, informs defensive actions and strategies to protect against malware threats. When adversaries launch cybersecurity attacks, tactics, techniques, and procedures (TTPs) are often reused, and their attacks are modelled based on specific strategies and techniques, leading to recognisable patterns even in the most sophisticated attacks. This has led the scientific community to model cybersecurity attack patterns and techniques from reported incidents to predict adversary behaviour and tactics \supercite{georgiadou2021assessing} \footref{conceptnote3}. The US-based research institute MITRE achieved a significant milestone by analysing the Tactics, Techniques, and Procedures (TTPs) used in real-world cyberattacks by Advanced Persistent Threat (APT) actors. It developed the ATT\&CK framework, a behavioural model \supercite{georgiadou2021assessing} that documents and classifies these TTPs based on several core components (Tactics, Techniques, Sub-Techniques, Procedures) and attack attribution \supercite{park2023destructive} \footref{conceptnote3}. The framework links specific TTPs to the particular APT groups that have employed them in real-world cyberattacks, enabling a better understanding of which groups are likely to utilise specific methods and facilitating the development of targeted defensive strategies.

The ATT\&CK framework does not represent a linear behaviour of an attacker's tactic or techniques in order from left to right, but may be employed in line with the motivations required to achieve operational goal \supercite{CISA2023BestPracticesMITRE} \footref{conceptnote3}. The ATT\&CK framework has successfully contributed to three domains: Enterprise, Industrial Control Systems (ICS), and Mobile. Some tactics are implemented through automation, while others require human execution. Certain tactics generate network traffic, whereas others do not; those that do may involve communication between the attacker and the victim's machine, or between systems within the victim's internal network \supercite{chen2023survey} \footref{conceptnote3}. 

Having outlined the core aspects of malware in the previous sections, the wider implications of malware can now be explored in the quantum domain. The following section therefore turns to a discussion of these implications in quantum technologies.

\section{Discussion: implications for the quantum era}
The chapter seeks to synthesise the findings of the review to explore the implications of malware behaviours in the quantum era by adopting the European Competence Framework for Quantum Technologies as a reference framework for investigating malware risks against quantum architecture ecosystems. The section also highlights key questions and considerations that need to be addressed in the critical quantum infrastructure domain.

The European Competence Framework for Quantum Technologies (CFQT) is a structured reference framework for analysing and understanding competencies needed in the quantum technology landscape to help advance quantum technology proficiency and guide qualifications, and job requirements. It is organised into several domains and subdomains and comprehensively captures the areas of knowledge and skills essential for the workforce of the quantum industry \supercite{cfqt2025, greinert2025extending}.

Although the CFQT is designed primarily to guide curriculum and competence development, it also offers a valuable lens for analysing the architectural complexity of quantum technologies in the context of cybersecurity. Specifically, it can be used to map the potential surfaces of malware attacks across various domains of quantum systems, highlighting where classical vulnerabilities intersect with emerging quantum infrastructure and applications.

This study uses CFQT to define quantum capabilities in computing, sensing, and communication, and to identify specific subsystems and interfaces that are susceptible to malware in the era of quantum technologies. The aim is to connect quantum domain knowledge with security awareness, creating a competence-informed perspective on threat modelling, particularly in hybrid quantum-classical systems.

\subsection{Impact on CFQT domains}
This section introduces two layers of analysis to understand malware in quantum environments. This first layer begins by applying a structural perspective to breakdown how malware can disrupt and impact each CFQT domain individually. Building on this, the next layer employs a cross-domain analysis in Table~\ref{tab:BehaviourMapping} by integrating the malware behaviours identified in the SLR with aspects of the CFQT. This enables an in-depth understanding of how malware operates and also provides insight into how malware can be strategically weaponised across interconnected systems such as the quantum Internet comprising composite systems that function holistically. Together, both analysis allows for a more abstracted, behaviour-driven risk modelling approach helping to form a dual-view analysis approach.

\subsubsection{CFQT-1: Concepts and foundations} 
\textbf{1.1 Basic quantum concepts; and 1.2 mathematical formalism and information theory}

Although this domain is theoretical, its integrity depends on unchanged training resources, learning materials, tools, and software, which can be vulnerable and subject to manipulation. Malware might find its way to poisoning the fundamental and foundational layer of quantum by sabotaging and poisoning simulators, Artificial Intelligence (AI) agents, or software tools that are hardcoded to simulate quantum mechanical behaviours or concepts, allowing researchers to publish flawed results or misinform educators about flawed assumptions of basic quantum principles, destabilising research efforts and academic progression. This would mean corrupting AI agents' training data or the simulator's core functions, physics models, so that every experiment produces biased/malicious outcomes. This is a very indirect malware approach and therefore subjective, as it does not directly compromise quantum architectures. A typical example is a research group that relies on an infected quantum simulator to model computations and validate results. However, without their knowledge, the simulator hides decoherence errors, fidelity degradation, and performance failures, resulting in the publication of a failed model if replicated on actual quantum hardware. Piazzesi et al. \supercite{piazzesi2021attack} demonstrate how a self-driving simulator such as CARLA can be compromised by injecting malicious changes into the simulated environment, such as altering detection labels, inserting fictitious obstacles, or altering sensor inputs that corrupt the training and testing data it produces. This work shows that poisoning the simulation environment inherently poisons all models developed from it.

\subsubsection{CFQT-2: Physical foundations of quantum technologies}
\textbf{2.1 Atomic physics; 2.2 quantum optics and electrodynamics; 2.3 solid-state physics; and 2.4 open quantum systems}

These theoretical principles form the foundation for understanding quantum technology engineering, directly underpinning the practical implementation involved in designing, building, and operating quantum hardware, protocols, and systems. These foundational layers are instrumental in guiding and shaping how quantum states interact within their environment. For quantum systems to reach their maximum operational fidelity, the preservation of quantum coherence \supercite{streltsov2015measuring, streltsov2017colloquium, xi2015quantum, divincenzo1999quantum} must be sustained; the stable phase relationship between quantum states. Decoherence \supercite{zurek2003decoherence, viola1998dynamical, schlosshauer2019quantum}, resulting from uncontrolled interactions between atomic particles and their surrounding environment, disrupts this delicate state or balance, leading to loss of quantum information. This phenomenon represents a fundamental obstruction towards the development of large-scale quantum architectures, as particles exhibiting quantum behaviour, such as superposition and entanglement, are intrinsically vulnerable to even minimal levels of environmental noise and perturbation. Malware designed to target and interact with a particle environment through its control systems \supercite{dong2010quantum, meshulach1998coherent, meshulach1999coherent, leonard2019digital, liebermann2016optimal} can introduce decoherence changing or collapsing its wave function. This can be achieved through low frequency noise\supercite{bergli2009decoherence}, pulse sequence \supercite{fraval2005dynamic} or control\supercite{xu2025security}, jitter\supercite{lao2023quantum}, time delays, signal interference, thermal noise, frequency modulations, and feedback errors. Such attacks on quantum architecture can lead to system downtime or distorted measurements and incorrect experimental results.

\subsubsection{CFQT-3: Enabling technologies and techniques}
\textbf{3.1 Laboratory techniques, noise and shielding; 3.2 solid-state technologies, nanotechnologies; 3.3 optical technologies; 3.4 control technologies; and 3.5 computers and software}

This domain explores the physical realisation of quantum technologies by employing laboratory techniques and enabling technologies needed to fabricate\supercite{cui2008nanofabrication}, scale, stabilise, integrate, and control \supercite{dong2010quantum} quantum technologies for practical computation, sensing, and communication tasks. Enabling technologies and techniques comprise practical technologies and engineering methods that allow the realisation and functionality of quantum systems. It deals with applied design and implementation by bridging the gap between theoretical foundations and the delivery of usable quantum solutions. This domain would be the most directly vulnerable to malware attacks as it contains the most classical hardware and software integrations, including experimental devices, cryogenic \supercite{hornibrook2015cryogenic}, control systems \supercite{hakelberg2018hybrid, zhang2023classical, oskin2002practical,}, optical devices \supercite{walmsley2015quantum}, noise, and shielding systems \supercite{malevannaya2025engineering}, mostly designed using classical implementations. Malware can distort, disrupt, or hijack the operations of quantum applications in real time by embedding itself into several layers of the classical stack, such as compilers, software, firmware, drivers, or even embedded in integrated circuits (hardware trojans), making attacks on quantum systems dangerously plausible today \supercite{das2023sok, ghosh2023primer}. Therefore, it is important to note that multiple layers of security are essential to ensure the resilience and integrity of even the most advanced quantum systems against complex threats and malicious behaviours.

\subsubsection{CFQT-4: Quantum hardware}
\textbf{4.1 Superconducting electronic circuits; 4.2 spin-based systems; 4.3 neutral atoms and ions; 4.4 photonic systems; 4.5 emerging qubit concepts; 4.6 quantum state control; 4.7 hybrid quantum-classical systems; and 4.8 technology realisation}

This section forms the essential foundations of quantum technologies as physical systems. By transforming quantum theory into practical engineering, physicists and engineers build coherent, high-fidelity functional physical quantum architectures and devices using solids at the nanometre scale or confined particles at the atomic scales, to advance and engineer quantum technologies. Examples include superconducting circuits\supercite{huang2020superconducting}, spin-based systems\supercite{kloeffel2013prospects, martinez2023information}, trapped ions\supercite{duan2010colloquium, bruzewicz2019trapped}, and photonic architectures\supercite{o2009photonic, flamini2018photonic, wang2020integrated}. 

Because the integration of classical and quantum hardware is the foundation of any functional quantum system, it is also a critical target for hardware-level threats. Adversaries may exploit vulnerabilities during the design, fabrication, or calibration phases of a system. Threats include the insertion of hardware Trojans\supercite{xiao2016hardware, karri2010trustworthy, bhunia2018hardware, becker2013stealthy,} in hybrid classical components that interface with quantum systems to cause malicious alterations to quantum state control mechanisms. Malicious trojans could be embedded in quantum hardware during the fabrication stage \supercite{das2023sok} by introducing deliberate alterations to the layout, circuitry, or manufacturing processes. Such sabotage can alter coherence, lower operational fidelity, induce unwanted crosstalk\supercite{sarovar2020detecting, ash2020analysis}, or create subtle, latent faults that remain hidden until the device is in active use.

Such attacks are severe because they occur early in the supply chain, before deployment, and remain dormant until triggered. These vulnerabilities not only risk system integrity, but can also affect experimental results, compromise scientific credibility, slow research progress, and damage investor or institutional trust. Protecting quantum hardware, therefore, requires rigorous physical security, trusted manufacturing processes, and multilayered verification protocols \supercite{beaumont2011hardware}.

\subsubsection{CFQT-5: Quantum computing and simulation} 
\textbf{5.1 Basics; 5.2 quantum simulators; 5.3 quantum programming tools and software stack, error correction; 5.4 quantum computing subroutines; 5.5 quantum algorithms; and 5.6 applications of quantum computing and simulation}

This domain focuses on applications of quantum technology used for computation \supercite{nielsen2010quantum, gambetta2017building}. Qubits, which represent the unit of information in quantum computing, can be applied in areas such as computation and simulation to solve complex problems or simulate natural phenomena\supercite{feynman2018simulating, georgescu2014quantum}. This builds on several logical abstractions, including compilers, transpilers, programming languages, libraries, and software stacks \supercite{Hidary2021, haner2018software, svore2006layered, bandic2022full, cross2017open, smith2016practical, shehata2025building, stirbu2023full, steiger2018projectq, larose2019overview, serrano2022quantum, hua2023qasmtrans,} to aid in the translation or provision of instructions that operationalise qubit states to perform computationally intensive tasks or simulations far too complex or inefficient for classical systems.

Malware can exploit vulnerability in software, libraries, or applications, creating opportunities for malicious injection operations and attacks on quantum circuits. For example, an extra gate that causes unwanted interference \supercite{ghosh2023primer}. Malicious actors can also embed backdoors in compilers, poison software, tools, extensions, or library updates, similar to a supply chain attack \supercite{martinez2021software, alkhadra2021solar, wolff2021navigating}, but specifically for the quantum software toolchain. This can be particularly dangerous for organisations relying on quantum computers for sensitive computations and simulations (e.g., military route and navigation, cryptography, or molecular simulation) as it could leak intermediate measurement data or compromise the logic, reasoning, and outcomes of quantum operations. These attacks distort quantum results and are difficult to detect especially when outcomes are inconclusive, given the inherently probabilistic nature of quantum outputs.

\subsubsection{CFQT-6: Quantum sensors and imaging systems}
\textbf{6.1 Basics; 6.2 electromagnetic field sensors; 6.3 temperature, particle and pressure sensors; 6.4 inertial and gravity sensors; 6.5 quantum imaging; 6.6 atomic clocks; and 6.7 applications of quantum sensors}

This section explores the application of quantum effects such as superposition and entanglement to detect, form images, and make precise measurements of physical quantities or aspects of the real world, such as changes in electromagnetic fields, gravitational waves, temperature, pressure, rotation, acceleration, frequency, and time, using extreme levels of precision and unparalleled accuracy \supercite{degen2017quantum,crawford2021quantum,dowling2015quantum,}. This domain showcases the wide-ranging applications of quantum sensors, some of which can potentially be integrated into critical infrastructure sectors, including battlefield threat detection, nuclear safeguards, high-precision timekeeping, earth observations, navigation in GPS-denied environments, and supporting advanced missile defence systems \supercite{krelina2021quantum, farley2021quantum, abraheem2025emerging, choi2023quantum, MalekosSmith2024QuantumPrimer}.

Similarly to quantum computing architectures, quantum sensors also depend on classical hardware, such as control systems, to control pulses or read out information \supercite{poggiali2018optimal, frey2017application, hincks2014accounting, liu2021hamiltonian, mabuchi2005principles,}, potentially creating a path for malware to compromise sensors confidentiality, integrity or availability. Malware can manipulate or distort control parameters, field sensors, pulse durations \supercite{goswami2003optical} and feedback control, to dampen entanglement\supercite{steane2014pulsed}, increase decoherence, or introduce noise, causing degraded performance or unreliable results. Malware could trigger false alarms or weaponise quantum sensors to hide real threats from enemy detection sensors creating slow reaction times during critical events such as nuclear attacks or seismic monitoring, which could lead to devastating consequences or to spoof quantum accelerometer or gyroscope data in navigation systems, to misguide aircraft, submarines, or autonomous vehicles, especially in GPS-denied environments. Calibration routines are essential to ensure the accuracy of sensors \supercite{d2004quantum}. Malware could shift baseline readings, making the sensor unreliable over time with no apparent signs of failure.

\subsubsection{CFQT-7: Quantum communications and networks}
\textbf{7.1 Basics; 7.2 quantum random number generators; 7.3 quantum key distribution; 7.4 applications of quantum cryptography; 7.5 infrastructure for quantum information networks (quantum internet); and system networks (composite systems), quantum internet applications}

This section explores the application of quantum information transmission and security across distances. By applying both theoretical and practical concepts such as the entanglement distribution \supercite{li2023entanglement, pompili2022experimental, shi2020concurrent}, quantum teleportation \supercite{bouwmeester1997experimental, pirandola2015advances, ma2012quantum,} and quantum channels \supercite{gyongyosi2018survey, caruso2014quantum}, the foundation is laid for the development of connected quantum systems \supercite{elliott2018darpa, wei2022towards, kozlowski2020designing, pompili2021realization}. The domains covers concepts such as Quantum Random Number Generators (QRNG)\supercite{ma2016quantum, herrero2017quantum, jennewein2000fast}, quantum key distribution protocol (QKD) to establish secure keys\supercite{cao2022evolution, scarani2009security, sasaki2011field, lo2005decoy}, the infrastructure of quantum networks such as repeaters\supercite{munro2015inside, azuma2023quantum, briegel1998quantum, azuma2015all, dur1999quantum, sangouard2011quantum, zwerger2012measurement}, optical fibres, satellites\supercite{bedington2017progress, liao2017satellite}, and control systems \supercite{diadamo2021distributed, valls2024brief, yamamoto2008avoiding, altafini2012modeling, petersen2014control} needed to scale communication and connectivity. Finally, it highlights the integration of quantum systems and devices such as computers, sensors, and communication nodes to form a global quantum Internet\supercite{wehner2018quantum, kimble2008quantum, azuma2023quantum}.

Quantum entanglement is the critical security mechanism of quantum communication used in QKD, teleportation, quantum networks \supercite{li2023entanglement, pompili2022experimental} and distributed computing. Quantum networks rely on classical controls and technological stacks to coordinate or route entanglements \supercite{shi2020concurrent}, schedule pulse signals, and manage errors. Malware may desynchronise timing signals between repeaters so that entanglement fails or tamper with entanglement-based communications by introducing subtle phase errors in entangled states, making the entanglement look intact but lacks correlations. Malware may also attempt to spoof a legitimate entangled partner to create a man-in-the-middle attack in an attempt to compromise key information. Malware may also aim to propagate across the quantum network via quantum repeaters, network updates similar to worms in distributed nodes \supercite{shoch1982worm} or make subtle corruption, calibration or configuration of data at multiple network points impacting availability of network \supercite{satoh2021attacking, shang2016quantum, amoretti2020entanglement, yu2009sudden, guo2017entanglement}.

\subsubsection{CFQT-8: Quantum communications and networks}
\textbf{8.1 Industry landscape and market analysis; 8.2 business strategy, entrepreneurship and management; 8.4 impact and responsibility; and 8.5 education and communication}

This domain moves from technical expertise to exploring how quantum technologies shape society, markets, and culture. It addresses how the quantum technology industry can increase its value through intellectual property, viable business models, strategic partnerships, ethical deployments, legal and regulatory frameworks, and building quantum literacy. This section examines the economic, policy, societal integration and responsible use of quantum technologies. Because this domain is people-centric rather than purely technical, malware impacts are indirect but deeply consequential. Malware may employ social engineering i.e. phishing\supercite{gupta2018defending} to gain confidential information such as calibration setups, intellectual property that shouldn't be shared publicly, creating layers of security and trust issues that inevitably compromise an organisations quantum advantage. Malware can also be used to gain access to critical quantum procedures, which might be used to disrupt business operations or supply chain schedules. Malware may also deliberately spread misinformation \supercite{sharevski2020beyond, sharevski2020tweet, zeng2020bad} using fear, false security claims, hoax, or data manipulation to exaggerate or downplay quantum capabilities, creating confusion, policy missteps, or erosion of trust toward an organisation's quantum integrity.

\begin{table}[H]
\vspace{-1cm}
\caption{Cross-domain analysis of malware behavioural threat mapping to CFQT domains}
\renewcommand{\arraystretch}{1.7}
\hspace*{-2.1cm} 
\begin{tabularx}{\dimexpr\textwidth+4cm\relax}{>
{\raggedright\arraybackslash}p{0.300\textwidth} >
{\raggedright\arraybackslash}p{0.350\textwidth} >
{\raggedright\arraybackslash}p{0.470\textwidth}} 
\toprule
\setlength{\leftskip}{1.5em}Malware behaviour & 
\setlength{\leftskip}{1em}CFQT domains &
\setlength{\leftskip}{1em}Quantum malware risk\\
\bottomrule

\setlength{\leftskip}{1.5em}Code organisation \supercite{chowdhury2022capturing} &
\setlength{\leftskip}{0em}3.5: computers and software; 4.7: hybrid quantum-classical systems; 5.3: quantum programming tools and software stack, error correction &
\setlength{\leftskip}{0em}Malware typically requires code to gain entry, initiate attacks, and exploit weaknesses within computational architectures \supercite{skoudis2004malware}. It leverages code structure to penetrate various layers of the software and hardware stack by taking advantage of insecure code or architectural flaws to exploit and embed backdoors that undermine system integrity. \\ 

\setlength{\leftskip}{1.5em}Evasion \supercite{veerappan2018taxonomy} &
\setlength{\leftskip}{0em}3.1: laboratory techniques, noise and shielding; 4.6: quantum state control; 6.7: applications of quantum sensors; 7.5: systems networks (composite systems), quantum internet applications & 
\setlength{\leftskip}{0em}To operate covertly and remain inconspicuous, quantum malware can hide within noise, manipulation of sensory instruments, recalibration of equipment data, logs\supercite{mitreT1070} or control flows, making detection extremely difficult due to quantum-classical measurement ambiguity. Through compromised repeater nodes, malware can route commands, use entangled states \supercite{satoh2021attacking} to mask the origin of an injected error, avoiding detection under the guise of legitimate network behaviour. \\

\setlength{\leftskip}{1.5em}Operations \supercite{sanchez2022toward} &
\setlength{\leftskip}{0em}3.5: computers and software; 4.6: quantum state control; 4.7: hybrid quantum-classical systems; 5.3: Quantum programming tools and software stack, error correction; 5.4: quantum computing subroutines; 5.5: quantum algorithms; 6.7: applications of quantum sensor; 7.5: Infrastructure for quantum information networks (quantum internet); 7.6: systems networks (composite systems), quantum internet applications &
\setlength{\leftskip}{0em}The success of any malware attack lies in a series of critical operations \supercite{mitreATTACK}. These actions are essential for achieving malicious goals such as breaching access controls, establishing persistent command channels, exfiltrating sensitive data, and ultimately disrupting or manipulating quantum computational processes and states. Depending on the malware characteristics, i.e. overarching objectives (complex, multi-stage goals), quantum malware may operate across both classical and quantum layers, leveraging multiple technical domains in a coordinated sequence or combination of attack stages, similar to a kill chain strategy. \\

\setlength{\leftskip}{1.5em}Dynamic \& memory features \supercite{maniriho2022BADTaxono} &
\setlength{\leftskip}{0em}3.5: computers and software; 4.7: hybrid quantum-classical systems & 
\setlength{\leftskip}{0em}Quantum malware can be triggered dynamically in a fileless manner on classical controls without dropping files to disk \supercite{sudhakar2020emerging} by influencing quantum states through observation and manipulation of classical registers that temporarily store quantum data. It can exploit memory regions (such as the heap or stack) to leak sensitive data like session tokens or cryptographic keys, or trigger buffer overflows \supercite{cowan1998stackguard, lhee2003buffer, deckard2005buffer} by corrupting adjacent memory. It may also inject malicious code into trusted processes or dynamically allocate memory within legitimate process address spaces to insert harmful payloads, potentially leading to a full system compromise and often evading even the most advanced security controls. Gaining access to memory spaces enables malware to exploit runtime behaviours such as feedback loops \supercite{liu2016comparing, wubshet2020investigation, james2011quantum}to manipulate quantum states or delay system recovery. \\

\end{tabularx}
\label{tab:BehaviourMapping}
\end{table}

\begin{table}[H]
\vspace{-1cm}
\caption*{{Table 5: }(\textit{continued)}}
\renewcommand{\arraystretch}{1.7}
\hspace*{-2.1cm} 
\begin{tabularx}{\dimexpr\textwidth+4cm\relax}{>
{\raggedright\arraybackslash}p{0.300\textwidth} >
{\raggedright\arraybackslash}p{0.350\textwidth} >
{\raggedright\arraybackslash}p{0.470\textwidth}} 
\toprule
\setlength{\leftskip}{1.5em}Malware Behaviour & 
\setlength{\leftskip}{1em}CFQT Domains &
\setlength{\leftskip}{1em}Quantum Malware Risk\\
\bottomrule

\setlength{\leftskip}{1.5em}Advanced Persistence \supercite{han2021aptmalinsight} &
\setlength{\leftskip}{0em}1: concepts and foundations; 2: physical foundations of quantum technologies; 3: enabling technologies and techniques; 4: quantum hardware; 5: quantum computing and simulation; 6: quantum sensors and imaging systems); 7: quantum communication and networks; 8: valorisation &
\setlength{\leftskip}{0em} An advanced persistent quantum malware will be typically employed by highly sophisticated threat actors or state sponsored groups who seek to gain unauthorised access to quantum architectures for a prolonged period \supercite{chen2014study, ussath2016advanced, alshamrani2019survey}. These types of malware attacks are known to be politically or economically motivated while leveraging a range of sophisticated TTPs that may not yet be documented in security databases, potentially involving techniques such as social engineering, evasion tactics, or physical breaches to disrupt operations, conduct espionage, or exfiltrate data. Such malware may progressively weaken different layers of the infrastructure, embedding itself over time, infiltrate learning tools or algorithms, deceiving scientists, establishing control and access, and chaining the compromised layers together to ultimately execute a devastating attack to weaken a state’s or organisation’s quantum ambitions whether scientifically, economically, or politically. \\

\setlength{\leftskip}{1.5em}Weaponisation \supercite{liles2015fusion} &
\setlength{\leftskip}{0em} 5: quantum computing and simulation; 6: quantum sensors and imaging systems; 7: quantum communication and networks &
\setlength{\leftskip}{0em}Once quantum malware has been deployed, its  infiltrated will seek to achieve objectives such as deceiving, disrupting, denying, degrading, or destroying. These objectives will be pursued against various quantum applications including computing, sensing and communication with potential consequences in societal, political, economic, or scientific domains.\\

\setlength{\leftskip}{1.5em}Deception \supercite{gorment2023machine} &
\setlength{\leftskip}{0em}1: concepts and foundations; 5: quantum computing and simulation; 6: quantum sensors and imaging; 7: quantum communication and networks; 8: valorisation &
\setlength{\leftskip}{0em} Apart from compromising quantum infrastructures directly, quantum malware may also seek to undermine trust across industry, business, ethics, and education through misinformation or poisoned educational material, simulators and learning kits. Malware might also be utilised to mislead funders, regulators, researchers or business managers to publish or push out bogus claims and products based on the malicious masking of defects or subtle errors such as environmental noise or decoherence fluctuations affecting quantum states which may have exaggerated performance results. Using social engineering, malware can be leveraged to deceive key stakeholders into revealing passwords, private keys or secrets that that facilitate authentication and authorization for accessing and operating critical quantum technologies. \\

\end{tabularx}
\end{table}

\subsection{Gaps that need to be addressed in the quantum era}

With rapid advancements in understanding quantum mechanics, the rise of quantum technologies and applications continues to accelerate. However, there is limited knowledge about how malware affects quantum computations, networks, and sensors. This highlights the need to expand the current understanding of malware threats beyond classical systems. Conducting such an investigation makes it possible to understand how malware compromises various components and behaviours within the quantum ecosystem, helping to identify relevant attack Tactics, Techniques, and Procedures (TTPs) to inform the development of appropriate defences for securing quantum technologies. Another essential line of investigation concerns the dependency of analogue control or classical platforms that interface with quantum hardware. These include software, artificial intelligence (AI) accelerators, arbitrary waveform generator (AWG), photonic control systems, cryogenic control systems, noise and shielding control systems, system-on-chip (SoCs), radio frequency system-on-chip (RFSoCs) and field-programmable gate arrays (FPGAs) that form quantum control apparatus for various quantum hardware.

Classical architectures remain essential for operating and modulating the signals required to sustain quantum state coherence, including entanglement and superposition, particularly within distributed quantum networks and processors. Examining the interdependence between quantum and classical architectures provides a critical foundation for understanding malware threats, as this interoperability of hardware is likely to generate new threats. Consequently, the scientific community, policymakers, and governments must collaborate to develop policies and defence systems that ensure vigilance and proactive security responses to these emerging threats.

Although not explored in depth within this paper, a promising direction for future research involves investigating the potential emergence of quantum-native malware, i.e. what does the future hold for quantum malware? Will entirely new forms of malware emerge that operate independently of classical counterparts, leveraging purely quantum operations and behaviours to compromise quantum networks and systems? Or will quantum malware evolve as a hybrid threat, integrating with classical malware techniques to infiltrate and disrupt quantum infrastructure? Key considerations include how such malware would behave, the distinctive features it might exhibit, the forms it could take, and the potential impact it could have, particularly in contrast to traditional malware models, especially when the presence of quantum networks or the Internet proliferates. This naturally raises various questions, including the necessary policies required for the identification, detection, and prevention of quantum malware attacks. In particular, if current security tools, policies, guidelines, and frameworks are adequate to protect against these types of threat, and if not, what tools, policies, or frameworks are needed to counter this type of threat?

As with any technology designed to improve the fundamental nature of human existence, looming threats such as malware will continually seek to undermine technological trustworthiness in unsuspected ways. This underscores the need to integrate security within the quantum technology infrastructure, as malware threats have the potential to remain unknown, alter desired outcomes, and hinder collective progress, particularly in critical moments when reliable performance is essential.

\section*{Acknowledgements}
The authors thank the University of South Australia for institutional support and access to facilities. Special thanks are extended to colleagues and supervisors whose constructive feedback has greatly improved the quality of this paper.

\section*{Funding}
This research was supported by the South Australian Government, Department of Industry, Innovation and Science, through the Industry Doctoral Training Centre (IDTC) PhD+ Program – Quantum Technology Applications.

\section*{Conflicts of interest}
The authors declare no conflict of interest.

\printbibliography{}
\end{document}